\documentclass[conference]{IEEEtran}

\newif\ifincluded
\includedfalse

\ifincluded
\fi

\usepackage[flushleft]{threeparttable}

\usepackage{listings}

\ifCLASSOPTIONcompsoc
  \usepackage[nocompress]{cite}
\else
  \usepackage{cite}
\fi

\usepackage{amsfonts}  

\usepackage{amsmath}  

\usepackage{multirow}

\usepackage{hyperref}

\usepackage{graphicx}

\begin{document}
\bstctlcite{IEEEexample:BSTcontrol}

%
\title{Atomic Crosschain Transactions for Ethereum Private Sidechains}


\author{
    \IEEEauthorblockN{
    	Peter Robinson\IEEEauthorrefmark{1}\IEEEauthorrefmark{2}\\ 
    	David Hyland-Wood\IEEEauthorrefmark{1}\IEEEauthorrefmark{2}, 
    	Roberto Saltini\IEEEauthorrefmark{1},
    	Sandra Johnson\IEEEauthorrefmark{1}, 
    	John Brainard\IEEEauthorrefmark{1}
         } 
    \IEEEauthorblockA{\IEEEauthorrefmark{1}Protocol Engineering Group and Systems (PegaSys), ConsenSys}
    \IEEEauthorblockA{\IEEEauthorrefmark{2}School of Information Technology and Electrical Engineering, University of Queensland, Australia}
    \IEEEauthorblockA{
    	peter.robinson@consensys.net, 
    	david.wood@consensys.net\\
    	roberto.saltini@consensys.net,
    	sandra.johnson@consensys.net,
    	john.brainard@consensys.net
	}
}

\maketitle

\thispagestyle{plain}
\pagestyle{plain}


\begin{abstract}
Public blockchains such as Ethereum and Bitcoin do not give enterprises the privacy they need for many of their business processes. Consequently consortiums are exploring private blockchains to keep their membership and transactions private. Ethereum Private Sidechains is a private blockchain technology which allows many blockchains to be operated in parallel. Communication is needed between Ethereum Private Sidechains to allow a function in a contract on one sidechain to execute function calls which return values from, or update the state of, another sidechain. We propose a crosschain technique which allows transactions to be executed atomically across sidechains, introduce a new mechanism for proving values across sidechains, describe a transaction locking mechanism which works in the context of blockchain to enable atomic transactions, and a methodology for providing a global time-out across sidechains. We outline the programming model to be used with this technology and provide as an example, a variable amount atomic swap contract for exchanging value between sidechains. Although this paper presents Atomic Crosschain Transaction technology in the context of Ethereum Private Sidechains, we discuss how this technology can be readily applied to many blockchain systems to provide cross-blockchain transactions.
\end{abstract}


%
\IEEEpeerreviewmaketitle

\section{Introduction}
In this paper we focus on Atomic Crosschain Transactions for Ethereum Private Sidechains \cite{robinson2018a}. Atomic Crosschain Transactions are motivated by two requirements common in other distributed systems: data and functionality we wish to use may be available in other systems. The first requirement, accessing data in other systems, has been previously explored by distributed query languages, for example SPARQL Federated Query 1.1 \cite{sparql-w3} and the Resource Description Framework 1.1 \cite{rdf-w3}. The second requirement, accessing functionality in other systems, has been common for decades via Remote Procedure Calls (RPC) \cite{rpc}.

Sidechains are blockchains which rely on a separate blockchain for their overall utility, such as enhanced security by \textit{pinning} to the blockchain \cite{robinson2019a}, for addressing information \cite{robinson2018b}, or for storing data which is used across all sidechains. We use blockchains as a shared data store for all sidechains participating in Atomic Crosschain Transactions.

Ethereum Private Sidechains \cite{robinson2018a} are ephemeral, on-demand, private, permissioned sidechains which provide confidentiality. Only the \textit{privacy} and \textit{permissioning} aspects of the technology are relevant to this paper. Privacy relates to keeping the identity of sidechain participants secret. Permissioning relates to restricting which nodes can connect to a sidechain and which Ethereum Accounts can be used with the sidechain.

Ethereum Transactions update the state of the distributed ledger of an Ethereum blockchain but can not return a value. Ethereum Views return values but can not update the state. In this paper we describe Crosschain Transactions that allow reading and writing across sidechains by combining Ethereum Transactions and Views in the following way: A Crosschain Transaction consists of an Originating Transaction and one or more Subordinate Transactions and Subordinate Views, where the Originating Transaction is the Ethereum Transaction which executes on the sidechain on which the Crosschain Transaction was submitted, and the Subordinate Transactions and Subordinate Views are Ethereum Transactions and Ethereum Views which execute on other sidechains as a result of the Originating Transaction. 

Reed \cite{reed1983} defines Atomic Actions as
\begin{quote}
...a program-specified computation that, although composed of primitive computational steps executed at different times and in different places, cannot be decomposed from the point of view of computations outside the atomic action. During the execution of atomic actions, intermediate states of data objects that arise will never be observed by computations outside the atomic action.
\end{quote}
In the context of Crosschain Transactions, \textit{atomic} means that the Originating Transaction and all Subordinate Transactions are either all accepted or all ignored. Enquiries as to the state of the distributed ledger on any sidechain after the Crosschain Transaction has started and before it has ended will yield a consistent value. Depending on the context of the read, the value returned will be the value prior to the start of the transaction, the value assuming the transaction is committed, or the read will fail.

A key consideration which previous distributed systems approaches did not need to overcome is \textit{consensus}. Blockchain systems require validator nodes to agree on the transactions to be added to the blockchain using a \textit{consensus} algorithm. Once the blocks are added to the blockchain, the updates are final. To deliver atomic behaviour across sidechains, we propose that updates once added to blockchains be considered \textit{provisional}. The \textit{provisional} updates need to be unrevocable. Once certain conditions occur, the updates need to be either committed or ignored.

\begin{figure}
  \includegraphics[width=\linewidth]{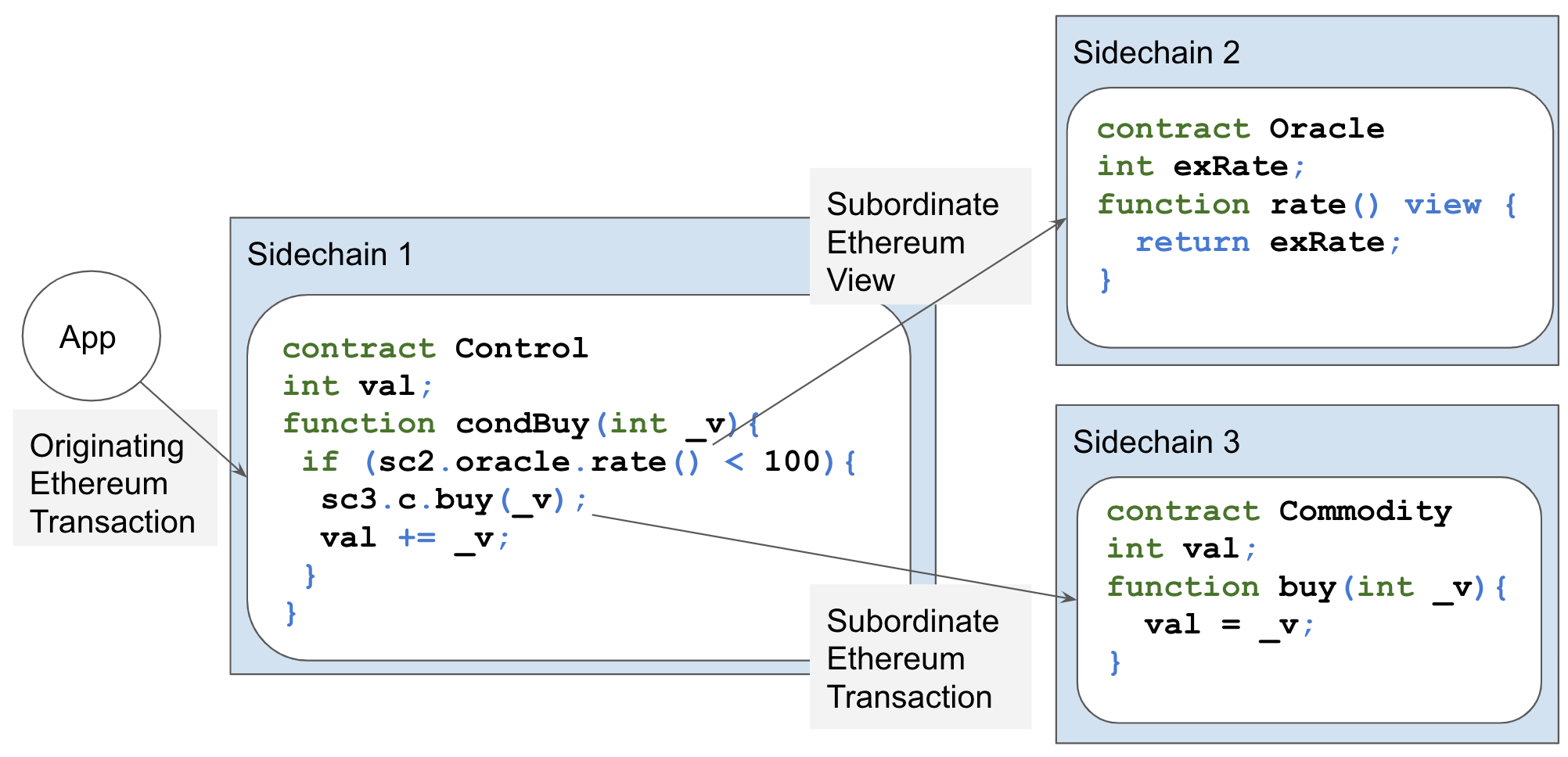}
  \caption{Crosschain Transaction}
  \label{fig:example}
\end{figure}

An example Crosschain Transaction consisting of an Originating Transaction, a Subordinate View and a Subordinate Transaction is shown in Figure \ref{fig:example}. The application submits the Originating Transaction to \texttt{Sidechain 1} which causes the function \texttt{condBuy} in the \texttt{Control} contract to execute. An Ethereum View call is dispatched to the \texttt{rate} function in the \texttt{Oracle} contract on \texttt{Sidechain 2}. The value returned by the \texttt{rate} function is used in the \texttt{condBuy} function. If the value is below \texttt{100}, then an Ethereum Transaction is dispatched to the \texttt{Commodity} contract on \texttt{Sidechain 3}. The \texttt{buy} function updates the state of the \texttt{Commodity} contract.

A more complex example could have multiple levels of Subordinate Transactions being submitted to different sidechains as a result of the Originating Transaction. Under the Originating Transaction and each Subordinate Transaction there could be multiple levels of Subordinate Views, calling out to different sidechains. The Subordinate Transactions and Subordinate Views that are issued are dependant on the call graph starting from the Originating Transaction.

We organised the paper as follows: in the \textit{Background} section we describe the concept of private blockchains and the enterprise version of Ethereum, explain the importance of having \textit{finality} for blocks so that they can be added to the blockchain, describe some key research related to distributed systems and two phase commits, and then explain threshold signature schemes. The next section, \textit{Related Works}, reviews alternative existing techniques for crosschain transactions, showing that they are not appropriate for atomic function calls across blockchains, and reviews existing work on blockchain locking. Thereafter, the proposed approach to Crosschain function calls is described in detail in the \textit{Approach} section, by first introducing the protocol components, then describing how Subordinate Views and Subordinate Transactions are processed, and finally describing the overall Crosschain Transaction and Crosschain View processing approaches. The \textit{Programming Model} to be used with the approach is described, including an example demonstrating how to use the approach to provide an atomic swap for partial amounts of Ether between sidechains. How the approach and programming model could fail are then analysed in two sections. \textit{Failure Cases Handled Within Protocol} explains how the protocol handles system failures and attacks. \textit{Failure Cases Handled by Application} describes failure situations which should be prevented by the application. The paper concludes with a discussion of how the approach could be applied to other blockchain systems to provide atomic cross-blockchain transactions.

This paper introduces what we believe is the first methodology for processing atomic transactions which include function calls across consortium sidechains. It describes how a threshold signature scheme can be used for proving values agreed on one sidechain to another sidechain. It introduces a method of parameter passing between sidechains which links the function doing the call in one sidechain to the function being called in another sidechain. It describes an atomic transaction locking mechanism which works in the context of blockchain. It introduces the idea of using a coordination blockchain as a global time-out, thus removing the requirement for crosschain time-out synchronisation. It introduces a crosschain transaction format and explains the methodology for processing the transactions within an Ethereum Virtual Machine. This paper contributes the concept of a Multichain Node. Multichain Nodes contain a set of trusted sidechain nodes all of which are on different sidechains, but work together. This paper introduces the programming model to be used for Crosschain Transactions.

\section{Background}
\subsection{Ethereum}
Ethereum \cite{wood2016a} is a blockchain platform which allows users to upload and execute computer programs known as Smart Contracts. Ethereum Smart Contracts can be written in a variety of Turing complete languages, the most popular being Solidity \cite{solidity}. Code is compiled into a bytecode representation. The bytecode can then be deployed using a contract creation transaction. Contracts have a special \textit{constructor} function which only runs when the contract creation transaction is being processed. This function is used to initialize memory and call other contract code. Miners execute the bytecode inside the Ethereum Virtual Machine (EVM). At present, each miner must execute all transactions for all contracts and hold the current value of all the memory associated with all of the contracts. The Ethereum community is actively working on methodologies to scale the Ethereum network by sharding the blockchain \cite{ethereum-sharding}.

Ethereum transactions update the state of the distributed ledger but do not return values. They fall into three categories: Ether transfer, contract creation, and calling a function on a contract. Ether transfer transactions move Ether from the user's account to another account. Contract creation transactions put code into the distributed ledger and call the constructor of the contract code, setting the contract data's initial state. Function call transactions call a function on a contract and result in updated state. Contract creation and function call transactions also allow Ether to be transferred. All types of transactions must be signed by a private key corresponding to an account and include a nonce value which prevents replay attacks. In addition to Ethereum transactions, ``View" function calls can be executed on the Smart Contract code. These View function calls return a value and do not update the state of the Smart Contract.

Precompiled Contracts are code which is accessible to Smart Contracts within the EVM at certain well-known contract addresses. Precompiled Contracts are implemented as part of the Ethereum Client in the Ethereum Client's own language. In the case of Pantheon Ethereum Client \cite{pantheon-github}, they are written in Java. Precompiled Contracts allow complex extensions to be added to Ethereum which would either take very long to execute or require resources outside of the EVM. As Precompiled Contracts are called from traditional Smart Contracts, to allow all nodes of an Ethereum network to execute all transactions, all of the nodes need to be instances of Ethereum Client software which supports the same set of Precompiled Contracts.

Executing code and accessing resources, such as memory, costs certain amounts of ``Gas". The ``Gas Cost'' of executing code is closely tied to the real world cost of executing each type of instruction. The current ``Gas Price'' is set for each block in terms of Ether by the miner who mines the block. Accounts instigating transactions specify the gas price they are prepared to pay for their transaction and specify the maximum amount of gas a transaction can use known as ``Start Gas". This commits an account holder to paying up to a certain amount of Ether for the transaction. Any unused gas is returned to the account holder at the end of the transaction. Miners reject transactions which run out of gas prior to completing execution.

In the Ethereum public network, ``MainNet", all contract code and data are readable by any user of any node which connects to the network. Smart Contracts on Ethereum MainNet can only perform permissioning in contract code, limiting which accounts can update the state of a contract. However, there is no mechanism to limit which users can read contract code and data. 

The value proposition of Ethereum is that it allows untrusted parties to use Smart Contracts hosted on a public, distributed, highly available, secure platform. 

\subsection{Private Blockchains and Enterprise Ethereum}
Private blockchains are blockchain networks which are established between nodes operated by enterprises \cite{robinson2018a}. Only permissioned nodes belonging to participating enterprises are allowed to join the private blockchain's peer-to-peer network and only permissioned accounts belonging to participating enterprises are allowed to submit transactions to the nodes. These blockchains provide the privacy and permissioning required by enterprises \cite{enteth20}. 

The need for security and permissioning features over and above what is available in standard Ethereum \cite{enteth20} has led to a range of platforms being developed. J.P. Morgan developed Quorum \cite{quorum-source}, a fork of the Golang Ethereum implementation called Geth  \cite{geth-github}. ConsenSys's Protocol Engineering Group, PegaSys created Pantheon \cite{pantheon-github}, an Ethereum MainNet compatible client which aims to meet the permissioning and privacy requirements of the Enterprise Ethereum Client Specification \cite{enteth20}. Hyperledger Fabric \cite{androulaki2018} is a distributed ledger platform originally created by IBM and now hosted by The Linux Foundation. Similar to Quorum and Pantheon, the platform offers privacy and permissioning features. Whereas Quorum offers Ethereum based private transactions, Pantheon offers private smart contracts which are private to a set of participants. Hyperledger Fabric offers the ability to host one or more smart contracts on a private blockchain called a ``channel". Hyperledger Fabric allows multiple channels to be operated on the one network, thus allowing for multiple sets of private contracts between different sets of participants to operate on the one network. An analysis of the merits of Hyperledger Fabric and Quorum can be found in \textit{Requirements for Ethereum Private Sidechains} \cite{robinson2018a}.

\subsection{Finality}
A block is deemed final when it can no longer be changed. In some consensus algorithms, such as PoW, finality is probabilistic, where as more blocks are added to the end of the blockchain, older blocks are less likely to be reordered. Consensus algorithms such as Istanbul Fault Byzantine Tolerant (IBFT) \cite{ibft} and Istanbul Fault Byzantine Tolerant version 2 (IBFT2) \cite{ibft2} give ``instant" finality, where once a transaction has been included in a block minted by a validator, it can no longer be changed.

\subsection{Distributed Systems and Databases}
Gray \cite{gray1978} and Lampson and Sturgis \cite{lampson1979} separately proposed two phase commit schemes which allow decentralised atomic updates. The first phase records a set of \textit{intentions} which indicate the data updates to be applied. The end of this phase is to request the transaction be committed. The second phase actually performs the update. If the second phase did not complete then it is reapplied as many times as needed to complete the algorithm.

Reed \cite{reed1983} proposed a methodology for processing atomic actions on decentralized data when faced with system failures. A feature of this methodology is that if a communications failure causes the second phase of a two phase commit to not reach a node, then when the node needs to access the data, it contacts other nodes to determine if the second phase occurred and the data should be committed or not.

\subsection{Threshold Signature Schemes}
Threshold cryptography schemes \cite{shamir1979} split secrets into Secret Shares. When a subset of participants cooperate they can reassemble the secret by combining their shares. In particular, any \texttt{M} shares of the \texttt{N} total shares can be used to recreate the secret. An attacker who has access to fewer than \texttt{M} shares is not able to determine any information about the secret. 

In the context of Threshold Signature Schemes the private key is the secret which is split. Any \texttt{M} shares of the \texttt{N} total shares need to be used to generate a signature. The \texttt{M} shares can not be brought together to reassemble the private key, as this would reveal the private key. Instead, the  \texttt{M} private key shares need to sign the data to be signed to generate \texttt{M} signature shares. These \texttt{M} signature shares are combined to create the threshold signature. The threshold signature can be verified using the public key which matches the private key which could be reassembled using the private key shares.

In a simple threshold signature scheme, the key shares are generated by a trusted party, called the dealer, then distributed to the participants. In a decentralised application, such a trusted setup is undesirable. An aggregated threshold scheme allows each participant to perform the operations of the dealer and create a set of key shares. These shares are then aggregated with the shares produced by the other participants to create the final key shares. This allows signing and verification to be done in a distributed fashion, without any trusted third party.

To ensure dealers are distributing correct values, the key generation scheme should be verifiable. In such a scheme, the final key shares are computed using an algorithm dependent on the private key shares of each participant. The Pederson commitment scheme \cite{ped1991} is an example of this form of verifiable secret sharing. The use of a verifiable generator proves that each participant is in possession of the private key share corresponding to its public share and makes it impossible for rogue participants to corrupt the process by submitting invalid shares. 

Aggregation of keys and signatures is only possible in signature schemes with special mathematical properties. In particular, signatures based on elliptic curve pairing such as the Boneh-Lynn-Shacham scheme \cite{bls2004}, support aggregation due to the bilinear property of the pairing operation shown in Equation \ref{equation:bls}. 
\begin{equation}
\label{equation:bls}
e(k*P, Q) = e(P, k*Q)
\end{equation}
	
This property allows signatures to be combined arithmetically, then verified by combining the corresponding public keys in a similar fashion. In blockchain applications, this allows all the signatures on all of the transactions in a block to be combined into a single signature which can be verified with the combined public key. 
 
Another advantage of using a BLS-based signature scheme is that it allows use of the \texttt{alt-bn128} curve \cite{nns2010}.This allows threshold signatures to be verified in the EVM as the EVM supports instructions to do this. The availability of EVM support makes on-chain operations significantly more efficient.

\section{Related Works}
\subsection{Overview}
Karynamoorthy et al. \cite{karunamoorthy2018} identified two core problems crosschain transactions need to overcome: communications between chains and the establishment of trust. Establishing trust is a prerequisite to communications. Within an Ethereum context, communication needs to encompass Ethereum Transactions and Ethereum Views \cite{wood2016a}. An Ethereum View executes a function which returns a value but does not  update the blockchain state. This can be considered as \textit{reading} data between chains. An Ethereum Transaction executes function calls which updates the blockchain state, can transfer value, but is unable to return results. This can be considered as \textit{writing} between chains. Another type of communication, value transfer, differs from general transactions which allow arbitrary functions calls, and typically involve transferring Ether, the underlying currency of Ethereum, or an ERC20 token \cite{eip20}\cite{erc20-standard}. 

The literature review below provides a summary of the existing research into blockchain value transfer and function calls. A more detailed review can be found in our ``Sidechains and Interoperability" paper \cite{johnson2019a}.

\subsection{Value Transfer}
Hashed Timelock Contracts \cite{hashtimelock} have been put forward as a mechanism for inter-chain value transfer. Smart Contracts are created on two separate blockchains, for instance Ethereum MainNet and a sidechain. A participant who wants Ether on the sidechain in exchange for Ether on Ethereum MainNet posts a message digest commitment to a secret to both contracts and puts in escrow the Ether in the contract on Ethereum MainNet. Another participant who wants to exchange Ether on Ethereum MainNet for Ether on the sidechain similarly puts Ether in escrow in the contract on the sidechain and posts a message digest commitment to a secret to both contracts. Both participants reveal their secrets and can then access their Ether. This allows trustless transfer of Ether between Ethereum MainNet and the sidechain. Given a fixed total quantity of Ether on the sidechain, this would allow for \textit{Mass Conservation} between Ethereum MainNet and the sidechain, where no additional Ether is created or destroyed. Building on the concepts of Hashed Timelock Contracts, Thomas and Schwartz have proposed an Interledger Protocol \cite{interledger}. Additionally, the Dogecoin to Ethereum bridge \cite{dogethereum} uses this technology to allow for transfer of coins between Doge blockchain and Ethereum. A limitation of Hashed Timelock Contracts is that the transfer can only be for the entire amount. There is currently no way to offer to exchange only part of the originally staked Ether.

Pegged Sidechains \cite{pegs2014} proposes Bitcoins to be transferred between the Bitcoin blockchain and sidechains, to allow for increased transaction rate and experimentation. The solution relies on publishing a proof that a transaction to transfer Bitcoin was included in a block and that further blocks were produced based on that block, in the \textit{source} blockchain. If the hashing power of the source blockchain is significant, then it would be impossible for an attacker to produce forge blocks. The solution requires a 24 hour confirmation period to ensure enough blocks have been produced based on the block with the transfer to provide adequate security. Wood \cite{polkadot2016} contends that the sidechain hashing power is unlikely to be sufficient to ensure security. Consequently, Bitcoins can be securely transferred to the sidechain from the Bitcoin blockchain, but not back. Moreover, this proposal is not appropriate for enterprise solutions, since confirmation times of 24 hours are unworkable. Additionally, this solution is limited to value transfer and does not provide a general purpose solution.

Minimum Viable Plasma \cite{plasma-mvp} builds on the concept of Plasma's \cite{poon2017} delegate Ethereum blockchains. Plasma chain operators create a Plasma Smart Contract on Ethereum MainNet and hold value deposited in the contract on a separate Plasma chain as Unspent Transaction Output (UTXO) \cite{uxto} values in a binary Merkle tree ordered by transaction index. Transactions on the Plasma chain involve proving that an unspent output had not previously been spent. Blocks on the Plasma chain are pinned to Ethereum MainNet. Two key challenges of this approach are the size of the proofs and exiting the Plasma chain to recover funds on Ethereum MainNet. The latter is particularly challenging, involving a complex exit procedure which includes a seven day challenge period.

Loom \cite{loom} have created a solution based on Plasma Cash \cite{plasma-cash}, which builds on the Minimum Viable Plasma approach to allow for the exchange of non-fungible assets. Each token has an identifier that represents the token's location in a sparse Merkle Tree. To spend a block, a proof needs to be submitted showing when the token has been used. The main limitations of this solution are that it is specific to non-fungible assets, does not allow generic value transfer and cannot be extended to offer crosschain function calls.

Crosschain communication in Metronome involves a two-step approach:  obtaining a \textit{proof of exit} Merkle receipt when removing tokens from the source chain and then presenting this receipt to the target blockchain to claim the MET tokens\cite{Metronome2018,Dale2018, johnson2019a}.  Metronome's design includes having autonomous smart contracts with ownership functions locked down after the launch \cite{Metronome2018}. Moreover, their cryptocurrency token, MET, is touted to be the first that is not permanently tied to any particular blockchain, and could be secured to another blockchain \cite{Metronome2018}. Metronome is focused on the portability of MET tokens between chains, and do not have any detail or design for atomic crosschain function calls. Moreover the validation phases of the movement of tokens are still under development at the time of writing.

The Token Atomic Swap Technology (TAST) research project proposes atomic crosschain asset transfers with the requirement that these assets need to exist as tokens, such as ERC20 tokens on the blockchains, \textit{independent of the blockchain's native currency} \cite{johnson2019a}. It implements a \textit{claim-first} transaction as a manifestation of the \textit{proof of intent} of the sender to make a crosschain asset transfer. There are \textit{witness rewards} for attesting to the transaction and an algorithm is provided for a cryptographically verifiable proof of intent \cite{Borkowski2018d,Borkowski2018,Borkowski2018c}. In order to implement this crosschain proposal all wallet balances have to be on all participating blockchains, and transferable assets need to exist as tokens on the blockchains \cite{johnson2019a}. The need to have wallet balances on all chains may limit the appeal of this approach, and performing atomic crosschain function calls, especially if they are more than one level deep, are likely to add a level of complexity that would be challenging to implement.

\subsection{Function Calls}
Cosmos \cite{cosmos2016} proposes a multi-blockchain system in which there are blockchains called Zones exchange tokens via a central blockchain called a Hub. The Zones and the Hub use Tendermint \cite{tendermint2018} a type of Practical Byzantine Fault Tolerance \cite{pbft1999}\cite{pbft2002} algorithm, rather than the Nakamoto Consensus \cite{nakamoto2018} used by Bitcoin. The value transfer uses a similar approach to Pegged Sidechains, posting proofs that a transaction has been included in a block. Similar to the Pegged Sidechains proposal, the solution relies on the security of the Zones (in the Pegged Sidechains case, the sidechains). 

Polkadot \cite{polkadot2016} proposes a multi-blockchain network built on Substrate consisting of \textit{Relay Chains}, \textit{Parachains} and \textit{Bridges}. Relay Chains, as the name suggests, relay messages between Parachains. Parachains receive and process transactions. Consensus for the entire system is provided by the Relay Chain. There are two main roles that participants play in the Polkadot ecosystem: \textit{Collator} and \textit{Validator}. Collators collect transactions on Parachains, propose blocks and provide zero knowledge non-interactive proofs proving the transactions result in valid state changes to the Validators. Groups of Validators ratify Parachain blocks and publish them to the Parachain. The Validators seal the Parachain block headers to the Relay Chain. The Validators are randomly assigned to Parachains, with the assignment changing regularly. Validators use a Proof of Stake consensus algorithm. Supportive roles are performed by \textit{Nominators} and \textit{Fishermen}. Nominators provide funds to Validators they trust to execute the Proof of Stake consensus. Fishermen observe the Parachains and submit fraud proofs to Validators.  

Two forms of crosschain transactions exist in Polkadot: Cross-Parachain and Polkadot to and from external chains such as Ethereum\cite{polkadot2016}. With Cross-Parachain transactions, the transactions are identical to typical transaction from external accounts. The transactions are moved from the outward bound queue on one Parachain to the incoming queue on another Parachain. Transactions from Polkadot to Ethereum are achieved by submitting transactions to a special multi-signature Ethereum contract. Transactions from Ethereum to Polkadot are achieved by calling into a special Ethereum contract which writes an event to the Ethereum event log. This event is interpreted as the outward bound call. The Polkadot system is complex because of its underlying requirement to allow heterogeneous blockchains. Routing Cross-Parachains transactions via the Relay Chain is likely to result in a bottleneck limiting performance. All Parachains connected to a Relay Chain will 
use its consensus mechanism. If a different consensus mechanism is required, then Polkadot proposes to achieve this via a bridge between a `Main' and a `Side' chain, but this functionality is still under development.

Wang et. al. \cite{blockchain_router} proposed a blockchain router system for connecting heterogeneous blockchains via a routing blockchain, in a similar way to Polkadot and Cosmos. Multiple \textit{Connector} components monitor each sub-chain for crosschain transactions, which they then submit to the router chain. The researchers suggested that the Connectors come to an agreement on the crosschain transactions, though how this agreement occurs is not described in the paper. \textit{Validator} components mint blocks on the router blockchain using a PBFT consensus algorithm. \textit{Connectors} take transactions from the router chain which are destined for their sub-chain and submit them to their sub-chain. There is no description of how  the Connector to submit transactions to the sub-chain is chosen, or what is done to ensure valid transactions are submitted by the Connector to the sub-chain. The system is kept secure by fining malicious actors using a complex economic incentivisation scheme and components called \textit{Surveillants}. 

Kan et. al. \cite{Kan2018AMB} proposed a crosschain protocol for heterogeneous blockchains via a router blockchain as an intermediary between chains in a similar way to Cosmos and Polkadot. In this scheme a three phase commit is proposed. The system appears to rely on reliable communications to ensure the system does not result in one chain perceiving a transaction has succeeded and the other chain perceiving that the transaction has failed.

Cross-Shard Contract Yanking \cite{cross-shard-contract-yanking}\cite{cross-shard-spec} has been proposed as a method of crosschain function calls for Ethereum 2.0 shards \cite{ethereum2-beacon-chain}. In this technique, a special EVM instruction is used to move a contract and its state temporarily from a shard to the current shard. The action occurs updating the state. The updated state is then returned to the originating shard inside a receipt, possibly with a Merkle proof \cite{nakamoto2008} proving that the state update was correct. Buterin \cite{cross-shard-contract-yanking} acknowledges that for this system to work, the contract needs to have small state and only be used by one entity. How the system responds to failures is not well defined. In particular, the case of the state being updated and the receipt being generated, but not being committed to the originating shard does not appear to be handled. From a confidentiality  perspective, this approach is problematic. Users of the contract on the originating shard could be different to users on the current shard. The users on the current shard should not be able to see the state of the yanked contract. The contract may also contain information which reveals the membership of the originating shard, which is problematic from a privacy perspective.

A mechanism for cross-Hyperledger Fabric channel communications \cite{hyperledger-fabric-interchain} has been proposed. The methodology is not documented, may not be secure, and does not appear to be supported by Hyperledger Fabric. 

BTC Relay \cite{btc-relay} is a method for allowing users of Ethereum to confirm Bitcoin transactions. Though not a method of executing function calls across chains, this technology is interesting as it allows actions to occur on one blockchain based on another blockchain. \textit{Relayers} are compensated for posting Bitcoin block headers to a Smart Contract on Ethereum. Bitcoin transactions are confirmed by users submitting Merkle proofs showing that a transaction belonged to a certain block. BTC Relay relies on PoW mining difficulty for its security. Multiple active Relay nodes must be prepared to post the block header for each block. In this way, if one Relay node posts a block header of a fork of the chain, other Relay nodes can post the block header of the longest chain. Transactions can only be validated if the block header they relate to is on the longest chain and if at least six block headers have been posted on top of the block header that the transaction relates to \cite{btc-relay-source}. As attackers can not produce a longer chain than the main Bitcoin blockchain due to the mining difficulty, they are unable to confirm transactions based on a malicious fork. PoW is not an appropriate consensus algorithm for private blockchains as organisations do not wish to allocate resources to mining of blocks \cite{enteth20}. Given the reliance of BTC relay on PoW mining difficulty, it is inappropriate for private blockchains.

The Clearmatics Ion project provides a framework and tools to develop crosschain smart contracts so that they execute automatically if a verifiable state transition has occurred on another database or blockchain \cite{Clearmatics2018c,johnson2019a}. This is known as `continuous execution'. Example code to implement this functionality is available from the Ion GitHub repository \cite{Clearmatics2018, Clearmatics2018c}. The system works by having blocks posted from one chain to another by a party that asserts a state transition happened on one chain. The relevant transactions are then extracted and validated via the relevant Merkle Patricia trie hashes via contracts on the receiving chain. The initial set up is via a series of steps deploying contracts on the two systems to allow the two systems to become known to one another.

\subsection{Blockchain Locking}
The Cross-Shard Contract Yanking \cite{cross-shard-contract-yanking}\cite{cross-shard-spec} methodology described above can be viewed as providing a mechanism for locking a contract. As described above, this technique is not appropriate for consortium chains as the confidentiality of the shard data is not maintained.

Ethereum Researcher Max C \cite{cross-shard-locking-scheme} describes a two phase commit locking scheme. The scheme requires locks to be committed to shards on which data to be atomically updated resides, and Merkle Proofs proving the state update which includes the lock be submitted to the shard which will execute the transaction on the data. This system requires knowledge of the block hashes of the shards the data resides on, on the shard which will execute the transaction, to allow the Merkle Proofs to be verified. Block hashes of one sidechain will not be available on other sidechains in the proposed system. Additionally, there is the risk that publishing block hashes of a sidechain may reveal information about the sidechain, thus compromising the confidentiality of the other sidechain.

A method for resolving deadlocks in Cross-Shard Locking has been proposed \cite{cross-shard-locking-resolving-deadlock}. This technique requires all cross-shard transactions to have a start block number. When lock contention occurs, the cross-shard transaction with the earlier start block number gains the lock. The methodology does not handle the case when the two contending transactions have the same start block number. The idea of using block number as a proxy for time in contract locking will be used in this paper. 

The Cross-Shard Contract Yanking proposal \cite{cross-shard-contract-yanking} allows contracts which include a \texttt{move\_to\_shard(uint256 shard\_id)} function to be yanked between chains. Having this function is equivalent to indicating that the contract can be \textit{locked}, and other contracts which do not contain this function are \textit{nonlockable}. The concept of Lockable and Nonlockable contracts will be used in this paper.

\subsection{Summary}
Many of the existing solutions involve a single relay chain to transfer transactions between chains, which could become a bottle neck. This will limit the transaction rate of the crosschain communications technique. Most of the existing solutions do not protect the privacy of members of the communication blockchains. This means that the techniques are not appropriate for enterprise blockchain scenarios. Some of the existing techniques require all blockchains to have the same consensus algorithm. Allowing different consensus algorithms on different sidechains is likely to be advantageous. Some of the techniques require centralised trusted parties to ensure the operation of the system. Centralisation goes against the ethos of blockchain. None of the techniques offer atomic crosschain function calls, and hence do not offer the capabilities of the proposed Atomic Crosschain Transaction technology.

\section{Approach}
This section describes the Atomic Crosschain Transaction protocol. It starts out by describing the protocol components, then combines the protocol components and presents the overall approach. The section concludes with Crosschain Views, a special case where all calls are Views; that is there are no transactions.

\subsection{Protocol Components}

\subsubsection{Multichain Nodes}
A Multichain Node is a grouping of one or more sidechain nodes, where each node is on a different sidechain. The sidechain nodes operate together to allow Crosschain Transactions and Views. The Multichain Node on which the transaction is submitted must have Validator Nodes on all of the sidechains on which the Originating Transaction and Subordinate Transactions and Views take place, plus have access to a Coordination Blockchain. 

Coordination Blockchains are used to coordinate crosschain transactions. These blockchains could be Ethereum Private Sidechains, Ethereum MainNet, or a private blockchain. All nodes on all sidechains which will participate in a crosschain transaction need to be able to access the Coordination Blockchain.

\begin{figure}
  \includegraphics[width=\linewidth]{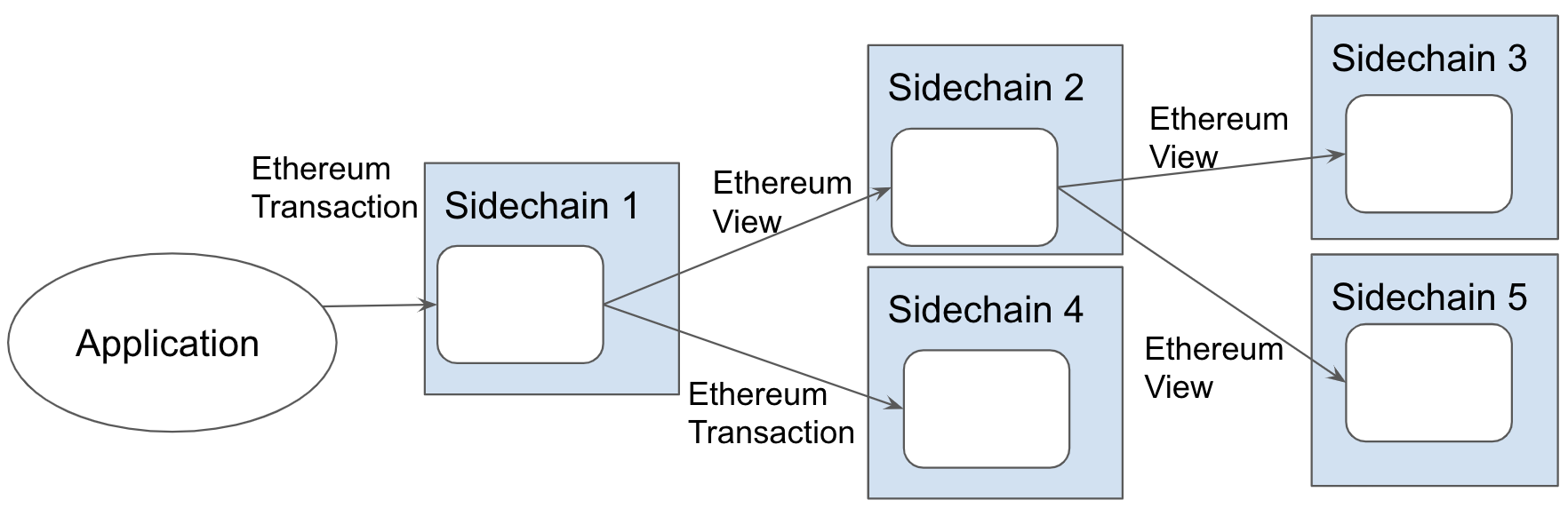}
  \caption{Crosschain Transaction Call Graph}
  \label{fig:callgraph}
\end{figure}

Consider the example shown in Figure \ref{fig:callgraph}. The application submits a transaction on Sidechain 1. This transaction causes an Ethereum View call on a contract on Sidechain 2 to execute, which results in other Ethereum View calls on Sidechain 3 and Sidechain 5. Additionally, as a result of the transaction on Sidechain 1, an Ethereum Transaction is submitted on Sidechain 4. This set of Ethereum Transactions and Ethereum Views is the result of a function call graph, in which a function on Sidechain 1 calls functions on Sidechain 2 to fetch a result, which in turn calls functions on Sidechain 3 and 5 to return results. The function on Sidechain 1 also calls a function on Sidechain 4.

Note that the result of submitting an Ethereum Transaction is to ask a blockchain to come to consensus on the result of the call. This differentiates blockchain activities from direct remote procedure calls (RPC) in traditional distributed systems.

\begin{figure}
  \includegraphics[width=\linewidth]{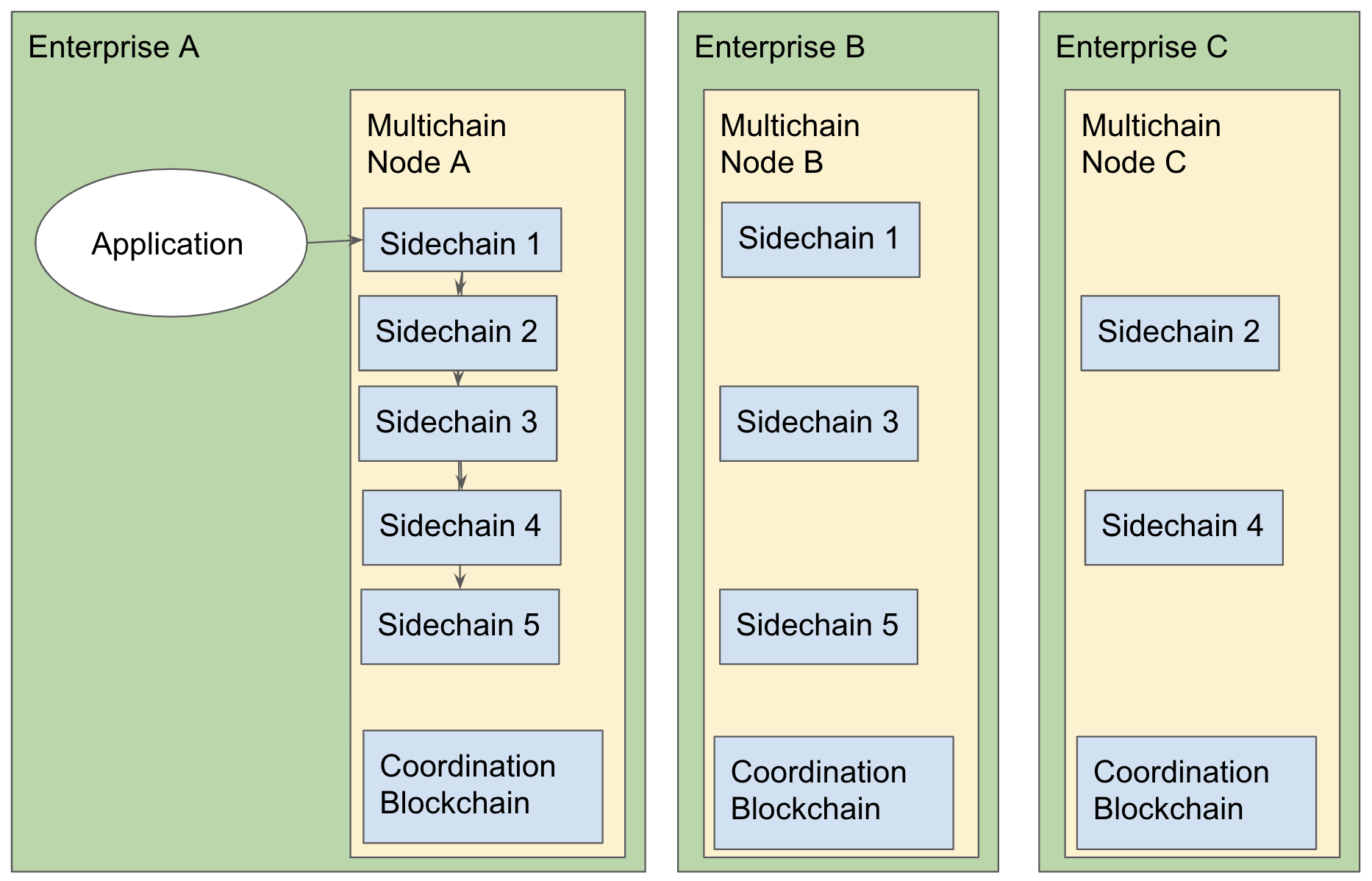}
  \caption{Enterprises, Multichain Nodes, and Sidechain Nodes}
  \label{fig:multichainnode}
\end{figure}

Figure \ref{fig:multichainnode} shows a set of enterprises which might take part in the call graph shown in Figure \ref{fig:callgraph}. Enterprise A operates Multichain Node A which contains validator nodes on each of the sidechains which make up the call graph. Enterprises B and C operate Multichain Nodes which do not have nodes on all of the sidechains. Enterprise B needs to be certain that the Ethereum Transaction on Sidechain 1 is final, despite not being able to access Sidechain 4. Enterprise C needs to be certain of the results returned by View calls on Sidechain 3 and 5. Enterprise C needs to be certain that the Ethereum Transaction on Sidechain 4 is final despite not being able to access Sidechain 1.

The sidechain on which the application submits the Originating Transaction is called the Originating Sidechain. For example, in Figure \ref{fig:callgraph}, it is Sidechain 1. The Originating Sidechain is transaction specific. It can be a different sidechain for each Crosschain Transaction. 

The sidechain nodes which are part of the Multichain Node on which the Originating Transaction was submitted are called Coordinating Nodes. The Coordinating Node for each sidechain is responsible for ensuring communications between sidechains. Each enterprise could submit Crosschain Transactions to their own Multichain Nodes, thus changing which nodes are deemed to be Coordinating Nodes each transaction.

\subsubsection{Sidechain Keys and Sidechain Threshold Signatures}
Messages from one sidechain can be verified as originating from the sidechain by use of a threshold signature scheme. Each validator node on each sidechain has a Sidechain Private Key Share. Any \texttt{M} of the \texttt{N} sidechain validator nodes must collaborate to sign a message. The Sidechain Public Key can be used to verify the signature. The signature and the public key do not betray any information about which nodes signed, what the threshold number of validator nodes is (\texttt{M}) , or what the total number of validator nodes on the sidechain are (\texttt{N}).

Assuming that a sidechain is using a Byzantine Fault Tolerant consensus protocol that offers finality and can cope with up to \texttt{F} validators being malicious, off-line, or faulty, then the threshold \texttt{M} should be set to \texttt{F+1}. This is because considering that at most \texttt{F} validators are malicious, if \texttt{F+1} validators agree on a given fact, for example that a transaction has been finalised, then this implies that at least one of these validators is honest and it sees the transaction as finalised in the sidechain. It should be noted that this threshold is less than the threshold required for consensus algorithms such as IBFT \cite{ibft} and IBFT2  \cite{ibft2} which are tolerant of \texttt{F} faulty nodes if \texttt{2F+1} nodes are not faulty.

Threshold private key generation occurs when the sidechain is established. The Sidechain Public Key needs to be published to the Coordination Blockchain. Any sidechain node can access the Sidechain Public Key once it is available in the Coordination Blockchain. When a validator node is added or removed from the sidechain, a new key generation must occur and the new public key must be published to the Coordination Blockchain. Publishing to the Coordination Blockchain involves a voting process between participants of the sidechain. The evaluation of the votes needs to reflect the threshold \texttt{M}. The voting process should be shielded such that the membership of the sidechain are not revealed. Salted hash shielding, similar to what has been used for Anonymous State Pinning for Private Blockchains \cite{robinson2019a} \cite{pinning_github} should be used.

\subsubsection{Sidechain Identifier}
Sidechain Identifiers are 256 bit values which identify a sidechain. They are used to identify which sidechain messages are intended for. They are also used to tie transactions to specific sidechains, to block replay attacks on other sidechains. Sidechain identifiers are randomly generated when the sidechain is first created as per the rules in Table \ref{table:sidechainidentifier}. The number range is chosen so as to not clash with the \textit{Chain Id} values used in Ethereum so that these blockchains can be specified using a Sidechain Identifier. A sidechain keeps the same Sidechain Identifier even if nodes are added or removed from the sidechain. 

\begin{table*}
  \centering
    \begin{tabular}{| l | l |}
    \hline
    Number Range & Description \\
       \hline
    0x00000000,00000000,00000000,00000000,00000000,00000000,00000000,00000000  &   Ethereum MainNet Chain Identifier  \\
    to & \\
    0x00000000,00000000,00000000,00000000,00000000,00000000,00000000,0000FFFF  &     \\
    \hline
    0x00000000,00000000,00000000,00000000,00000000,00000000,00000000,0000FFFF  &   Reserved for future use \\
    to & \\
    0xFEFFFFFF,FFFFFFFF,FFFFFFFF,FFFFFFFF,FFFFFFFF,FFFFFFFF,FFFFFFFF,FFFFFFFF  &     \\
    \hline
    0xFF0000000,00000000,00000000,00000000,00000000,00000000,00000000,00000000  &   Ethereum Private Sidechains  \\
    to & \\
    0xFFFFFFFF,FFFFFFFF,FFFFFFFF,FFFFFFFF,FFFFFFFF,FFFFFFFF,FFFFFFFF,FFFFFFFF  &     \\
    \hline
  \end{tabular}
  \caption{Sidechain Identifiers}
  \label{table:sidechainidentifier}
\end{table*}

\subsubsection{Crosschain Transaction Identifier \& Originating Sidechain Identifier}
The Sidechain Identifier on the Originating Sidechain is called the Originating Sidechain Identifier. When an application submits a Crosschain Transaction to the Coordinating Node on the Originating Sidechain, it randomly generates a Crosschain Transaction Identifier. The combination of the Crosschain Transaction Identifier, the Originating Sidechain Identifier, the Coordination Blockchain Identifier (see Section \ref{section:format}, \textit{Crosschain Transaction Format}), and the Crosschain Coordination Contract address (see Section \ref{section:format}, \textit{Crosschain Transaction Format}) globally identifies messages and transactions which belong to the Crosschain Transaction.

\subsubsection{Crosschain Coordination Contract}
Crosschain Coordination Contracts exist on Coordination Blockchains. They allow sidechain nodes to determine whether the state updates related to the Originating Transaction and Subordinate Transactions should be committed or not. The contract is used to determine a common time-out for all sidechains.

The contract contains a mapping between the message digest of the Crosschain Transaction Identifier and the Originating Sidechain Identifier, and Crosschain Transaction information. The information is:

\begin{itemize}
\item Transaction Timeout Block Number: This timeout is based on the Coordination Blockchain block number. If the Crosschain Transaction Commit message has not been posted to the Crosschain Coordination Contract prior to this block number, then the Crosschain Transaction is deemed to have timed-out. If the transaction times out, all provisional updates due to the Crosschain Transaction must be discarded. This value is calculated by the Crosschain Coordination Contract when it accepts a Crosschain Transaction Start message as the current block number plus the Crosschain Transaction Timeout value contained in the Crosschain Transaction Start message.
\item State Indicator:
 \begin{itemize}
 \item Started: Crosschain Transaction Start message was received and accepted by the Crosschain Coordination Contract. No further message has been received.
 \item Committed: Crosschain Transaction Commit message, indicating the Originating Transaction and all Subordinate Transactions state updates should be committed, was received and accepted by the Crosschain Coordination Contract. 
 \item Ignored: Crosschain Transaction Ignore message, indicating the Originating Transaction and all Subordinate Transactions state updates should be discarded, was received and accepted by the Crosschain Coordination Contract. This state is used to terminate Crosschain Transactions prior to the time-out expiring.
 \end{itemize}
\end{itemize}

The rationale for message digesting the Crosschain Transaction Identifier and the Originating Sidechain Identifier is to tie the Crosschain Transaction Identifier to the Originating Sidechain, without having to allocate any storage space to explicitly store the Originating Sidechain Identifier. 

A different Coordination Blockchain, and Crosschain Coordination Contract, can be used with each Crosschain Transaction. They are specified by the Coordination Blockchain Identifier and the Crosschain Coordination Contract address which are included in all Originating Transactions, Subordinate Transactions and Subordinate Views.

\subsubsection{Crosschain Transaction States}
\begin{figure}
  \includegraphics[width=\linewidth]{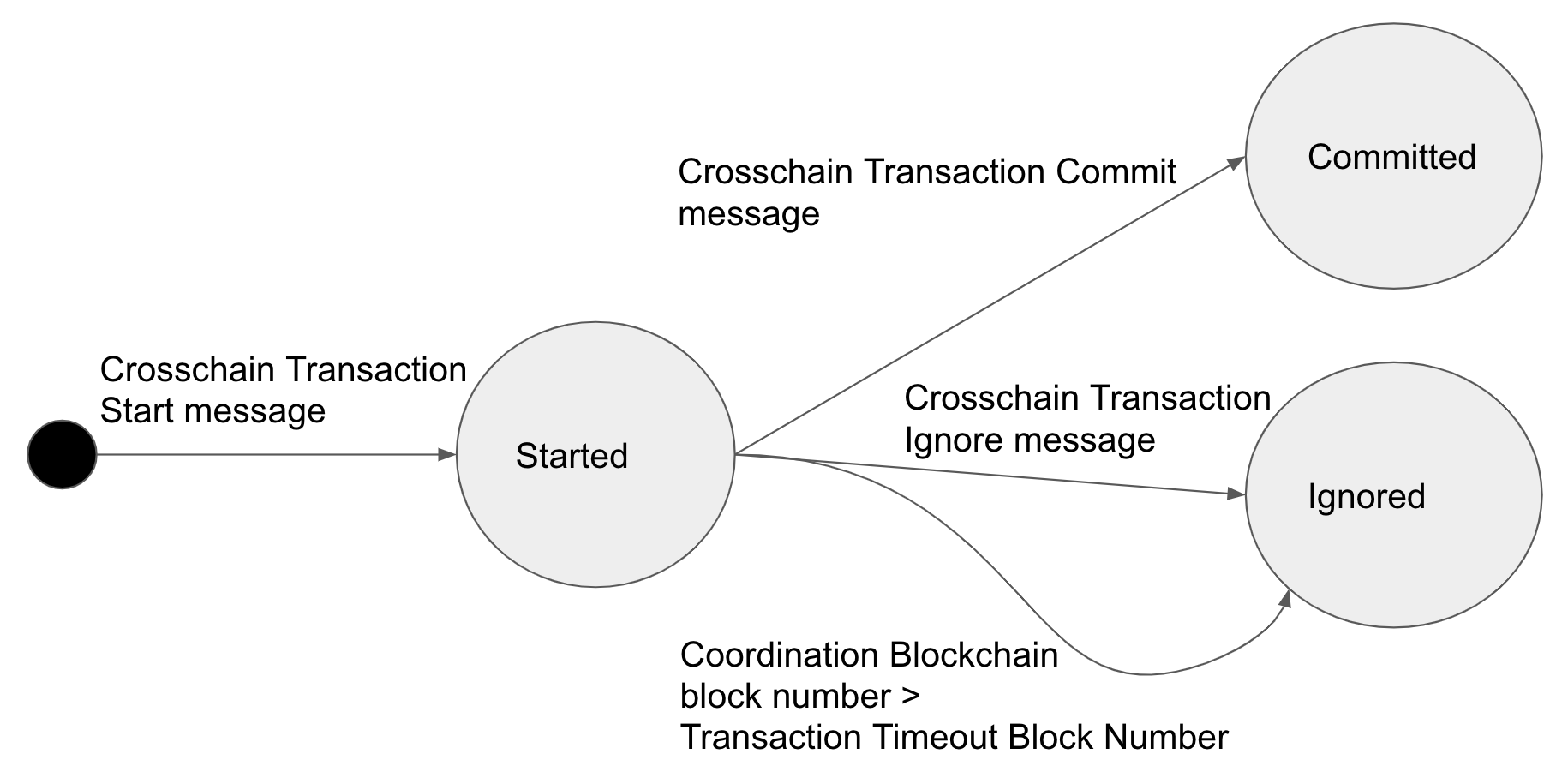}
  \caption{Crosschain Transaction States}
  \label{fig:transactionstates}
\end{figure}

Figure \ref{fig:transactionstates} shows how a Crosschain Transaction transitions between states. The state is held in the Crosschain Coordination Contract. Most of the state changes relate to messages described in detail in Section \ref{section:thresholdmessages}, \textit{Crosschain Threshold Messages}.

The Coordinating Node on the Originating Sidechain works with other Originating Sidechain validators to threshold sign a Crosschain Transaction Start message. This message contains the Crosschain Transaction Identifier, the Originating Sidechain Identifier, and the Crosschain Transaction Timeout. Validators will only sign this message if they agree with the proposed time-out, and are happy to process the proposed Crosschain Transaction. For example, a validator might choose to not sign the message if it felt that the Coordinating Node was spamming the network. The Coordinating Node on the Originating Sidechain submits the signed message to the Crosschain Coordination Contract. The Crosschain Coordination Contract accepts the Crosschain Transaction Start message if the Originating Sidechain's Sidechain Public Key can be used to verify the message. It creates a map entry for the Crosschain Transaction, sets the Transaction Timeout Block Number and sets the state to Started. 

In due course, the Crosschain Transaction is ready to be committed. The Coordinating Node on the Originating Sidechain works with other Originating Sidechain validators to threshold sign a Crosschain Transaction Commit message. The Coordinating Node on the Originating Sidechain submits the signed message to the Crosschain Coordination Contract. The Crosschain Coordination Contract accepts the Crosschain Transaction Commit message if the Originating Sidechain's Sidechain Public Key can be used to verify the message, and if the block number on the Coordination Blockchain is less than or equal to the Transaction Timeout Block Number for the transaction. 

To reduce resource usage, if the Coordinating Node on the Originating Sidechain determines that the Crosschain Transaction is going to fail, it should cancel the transaction. To cancel the transaction, the Coordinating Node on the Originating Sidechain works with other Originating Sidechain validators to threshold sign a Crosschain Transaction Ignore message. The Coordinating Node on the Originating Sidechain submits the signed message to the Crosschain Coordination Contract. The Crosschain Coordination Contract accepts the Crosschain Transaction Ignore message if the Originating Sidechain's Sidechain Public Key can be used to verify the message, and if the block number on the Coordination Blockchain is less than or equal to the Transaction Timeout Block Number for the transaction. 

The Crosschain Transaction times-out if neither a Crosschain Transaction Commit message or Crosschain Transaction Ignore message are submitted to the Crosschain Coordination Contract prior to the block number on the Coordination Contract being greater than the Transaction Timeout Block Number. At this point, all state updates related to the Crosschain Transaction can be ignored.

Publishing the transaction start, commit or ignore state along with the time-out to the Coordinating Chain allows all sidechains to use this chain as a global time-out clock and global state store. Acting as a global time-out clock, it means that each chain does not have to rely on its local understanding of time for time-outs, which would lead to race conditions in which one sidechain might commit a state update and another sidechain might ignore a state update. Acting as a global holder of state ensures sidechains which receive Subordinate Transactions can be sure that the requested transaction and associated time-out was approved by all validators on the Originating Sidechain.

\subsubsection{Contract Locking and Provisional State Updates}
When a Coordinating Node on a sidechain receives an Originating Transaction, Subordinate Transaction or View which is part of a Crosschain Transaction, it checks whether the contract is \textit{locked}. If the contract is locked, then the transaction or view fails. If the contract isn't locked, then the transaction or view can proceed. More complex behaviour is being considered for future work. However, to avoid complexities such as deadlocks, a simple \textit{fail if locked} approach is appropriate for this initial protocol version.

\begin{figure}
  \includegraphics[width=\linewidth]{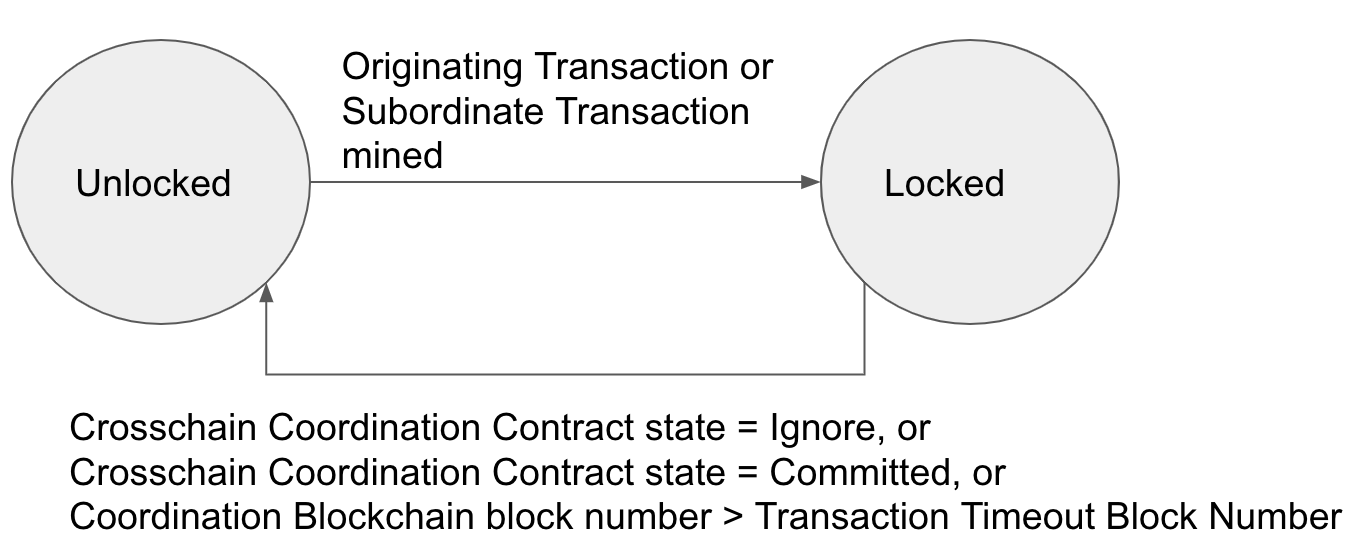}
  \caption{Contract Locking States}
  \label{fig:contractlockingstates}
\end{figure}

Figure \ref{fig:contractlockingstates} shows the locking state transitions for a contract. The Crosschain Coordination Contract will be in \textit{Started} state. The act of mining an Originating Transaction or Subordinate Transaction and including it in a blockchain locks the contract. The contract is unlocked when the Crosschain Coordination Contract is in the \textit{Committed} or \textit{Ignored} state, or when the block number on the Coordination Blockchain is greater than the Transaction Timeout Block Number. The Crosschain Coordination Contract will change from the \textit{Started} state to the \textit{Committed} state when a valid Crosschain Transaction Commit message is submitted to it, and it will change from the \textit{Started} state to the \textit{Ignored} state when a valid Crosschain Transaction Ignore message is submitted to it. 

Ordinarily, all nodes will receive a message indicating that they should check the Crosschain Coordination Contract when the contract can be unlocked. The message is sent from the Coordinating Node on the Originating Sidechain to the Coordinating Nodes on all sidechains involved in the Crosschain Transaction and from there to all nodes on all sidechains. When a node first processes a transaction, it will set a local timer which should expire when the Transaction Timeout Block Number is exceeded. If the node has not received the message by the time the local timer expires, the node checks the Crosschain Coordination Contract to see if the Transaction Timeout Block Number has been exceeded and whether the updates should be committed or ignored.

When a contract is first deployed it is marked as a Lockable Contract or a Nonlockable Contract. A Nonlockable Contract is one which can not be locked. This means that Originating Transactions or Subordinate Transactions will fail if they call the contract as the contract can not be locked. To match existing behaviour, the default when a contract is deployed is for the contract to be marked as a Nonlockable Contract. See the \textit{Programming Model} section, in Section \ref{section:programming-model} for an example of how Nonlockable and Lockable Contracts can be used together.

\subsubsection{Crosschain Transaction Format}
\label{section:format}
The Originating Transaction, Subordinate Transactions and Subordinate Views are all signed. In standard Ethereum, Views execute locally on a single node. The Subordinate Views in this system executes on all validators of a sidechain. Having the Subordinate Views signed should help prevent spamming.

Originating Transactions contain Subordinate Transactions and Subordinate Views in a tree-like structure, with Subordinate Transactions containing other Subordinate Transactions and Subordinate Views below them and Subordinate Views containing other Subordinate Views. The signature of an Originating Transaction, Subordinate Transaction or Subordinate View is across all contained Subordinate Transactions or Subordinate Views. Doing this ensures that nodes processing Subordinate Transactions and Views can be sure that the nested Subordinate Transactions and Views have not in some way being replayed or tampered with.

The Originating Transaction and all Subordinate Transactions and Views contain: 
\begin{itemize}
\item Type: Originating Transaction, Subordinate Transaction, or Subordinate View
\item Coordination Blockchain Identifier: A Sidechain Identifier which identifies the Coordination Blockchain to use for this transaction.
\item Crosschain Coordination Contract address: The address of the Crosschain Coordination Contract on the Coordination Blockchain to use for this transaction.
\item Crosschain Transaction Time-out: (Originating Transaction only): The length of time-out measured in Coordination Blockchain blocks.
\item Crosschain Transaction Identifier: Combined with the Originating Sidechain Identifier, Coordination Blockchain Identifier, and the Crosschain Coordination Contract address gives a globally unique reference to the transaction.
\item Originating Sidechain Identifier: Sidechain Identifier of the Originating Sidechain.
\item Sidechain Identifier: (Subordinate Transactions and Views only) Sidechain Identifier of the sidechain to execute this Subordinate Transaction or Subordinate View on.
\item Nonce: Standard Ethereum nonce which is unique per account per sidechain.
\item GasPrice: The amount offered to pay for gas for the transaction.
\item GasLimit: The maximum gas which can be used by the transaction.
\item To: Address of the account to send the value to, or the address of a contract to call.
\item Value: The amount of Ether to transfer.
\item Data: The RLP encoding of the truncated function signature hash and the function parameters.
\item Array of Subordinate Transactions and Subordinate Views which are called directly from this Originating Transaction, Subordinate Transaction or Subordinate View. Each array element could contain a nested tree of Subordinate Transactions and Subordinate Views.
\item V: part of the signature.
\item R: part of the signature.
\item S: part of the signature.
\end{itemize}
In EIP 155 \cite{eip155}, when the signature is calculated, the \texttt{V} value is combined with the Chain Identifier. As the Chain Identifier is included in the number range of the Sidechain Identifier, there is not need to combine the \texttt{V} value with the Chain Identifier.

\subsubsection{Function Call Processing}
When a node prepares to process an Originating Transaction, Subordinate Transaction or Subordinate View, it creates an ordered list of Subordinate Transactions and Subordinate Views it expects to execute. It uses these lists to ensure the code is executing as expected.

The high level code shown in Figure \ref{fig:example} is an abstraction of how Subordinate Transactions and Subordinate Views are called from within a function. Access to Subordinate Transactions and Subordinate Views is provided by precompiled contracts \cite{wood2016a}, one for Subordinate Transactions and one for Subordinate Views. The high level code is preprocessed to translate it from the high level syntax to precompile calls. 

When code is being processed, a call to a function in a contract on another sidechain results in one of the precompiles being executed. The precompile is passed the sidechain identifier of the sidechain the function should be executed on, the function identifier, and the parameters. The parameter values passed are the actual values that the EVM has for the variables at that point. The precompile compares actual values with the signed values for the next signed Subordinate Transaction or Subordinate View. The signed values are the values that the application expected to be passed in. If the actual values do not match the signed values of the next Subordinate Transaction or Subordinate View then the function call has failed. At this point, the entire Crosschain Transaction can be aborted.

\subsubsection{Crosschain Transaction Generation}
A Crosschain Transaction consists of the Originating Transaction and nested Subordinate Transactions and Subordinate Views. Each of these transactions and views contains signed parameters. These signed parameter values must match the actual values passed in via the EVM, as described in the previous section, \textit{Function Call Processing}. Further, the Subordinate Transactions and Views must be put into the Originating Transaction in the order in which they will execute. A Dynamic Program Analysis approach is recommended for determining the order of Subordinate Transactions and Views and the parameter values.

\subsubsection{Permissioning}
Enterprises may wish to restrict which accounts can execute Subordinate Transactions or Views on sidechains. This permissioning extends the existing Account Permissioning of Enterprise Ethereum \cite{enteth20} to include permissioning for Subordinate Views. Allowing enterprises to restrict which accounts can submit Subordinate Views is appropriate as they must be executed across all validators on a sidechain, and hence incur expense for all validators.

\subsubsection{Crosschain Threshold Messages}
\label{section:thresholdmessages}
The Crosschain Transaction protocol relies on threshold signed messages to prove to other sidechains that a sidechain has come to consensus on some information. The messages are threshold signed by the validators on the sidechain that wishes to prove the information. The messages can be verified by using the Sidechain Public Key available in the Crosschain Coordination Contract. Table \ref{table:thresholdmessages} lists the messages and their contents.

\begin{table*}
  \centering
    \begin{tabular}{| l | l | l |}
    \hline
    Message & Description & Contents \\
    \hline
    Crosschain Transaction Start  &   Submitted by the Coordinating Node on the Originating Sidechain to the & Originating Sidechain Identifier  \\
                                                    &   Crosschain Coordination Contract to indicate the start of a Crosschain    & Crosschain Transaction Identifier \\
                                                    &   Transaction.                                                                                                & Coordination Blockchain Identifier \\
                                                    &                                                                                                             & Crosschain Coordination Contract address \\
                                                    &                                                                                                             & Crosschain Transaction Time-out \\
                                                    &                                                                                                             & \hspace{3mm}measured in Coordination \\
                                                    &                                                                                                             & \hspace{3mm}Blockchain number of blocks. \\
    \hline
    Crosschain Transaction Commit &Submitted by the Coordinating Node on the Originating Sidechain to the & Originating Sidechain Identifier  \\
                                                    &   Crosschain Coordination Contract to indicate that the transaction should & Crosschain Transaction Identifier \\
                                                    &   be committed.                                                                                   & Coordination Blockchain Identifier \\
                                                    &                                                                                                             & Crosschain Coordination Contract address \\
    \hline
    Crosschain Transaction Ignore&Submitted by the Coordinating Node on the Originating Sidechain to the & Originating Sidechain Identifier  \\
                                                    &   Crosschain Coordination Contract to indicate that the transaction should & Crosschain Transaction Identifier \\
                                                    &   be ignored.                                                                                        & Coordination Blockchain Identifier \\
                                                    &                                                                                                             & Crosschain Coordination Contract address \\
    \hline
    Subordinate Transaction Ready& Sent by the Coordinating Node on the sidechain on which the Subordinate  & Originating Sidechain Identifier  \\
                                                    &  Transaction executed to the Coordinating Node on the Originating   & Crosschain Transaction Identifier \\
                                                    & Sidechain to indicate the Subordinate Transaction has been mined, the  & Coordination Blockchain Identifier \\
                                                    & transaction is final, and the updates are ready to be committed.       & Crosschain Coordination Contract address \\
                                                    &                     & Sidechain Identifier of the sidechain \\
                                                    &                      & \hspace{3mm}that the Subordinate Transaction  \\
                                                    &                                                                                                      & \hspace{3mm}executed on. \\
                                                    &                                                                                                      & Transaction Hash \\
    \hline
    Subordinate View Result        & Sent by the Coordinating Node on a sidechain which has executed a & Originating Sidechain Identifier  \\
                                                    & Subordinate View to the Coordinating Node on the sidechain which  & Crosschain Transaction Identifier \\
                                                    & called the Subordinate View to convey the result of the view.             & Coordination Blockchain Identifier \\
                                                    &                                                                                                             & Crosschain Coordination Contract address \\
                                                    &                                                                                                             & Sidechain Identifier of the sidechain \\
                                                    &                                                                                                             & \hspace{3mm}that the Subordinate View \\
                                                    &                                                                                                             & \hspace{3mm}executed on. \\
                                                    &                                                                                                             & Block number when the Subordinate \\
                                                    &                                                                                                             & \hspace{3mm}View executed. \\
                                                    &                                                                                                             & Subordinate View Hash \\
                                                    &                                                                                                             & Result \\
    \hline
  \end{tabular}
  \caption{Crosschain Threshold Messages}
  \label{table:thresholdmessages}
\end{table*}

\subsection{Subordinate View Processing}
\label{section:subordinateviewprocessing}
Subordinate Views can be submitted to a sidechain as a result of the Originating Transaction, Subordinate Transactions, or other Subordinate Views. This section presents how Subordinate Views should be processed.

In the sequence diagrams below `Sidechain B' is the sidechain on which the Subordinate View is processed. This sidechain has been named to differentiate it from the sidechain submitting the Subordinate View. 

The execution of Subordinate Views can be recursive. That is one Subordinate View can call another.

\subsubsection{Subordinate View Processing: Coordinating Node}
Together figures \ref{fig:subordinateviewcoord1} and \ref{fig:subordinateviewcoord2} show sequence diagrams for the processing a Subordinate View from the perspective of a Coordinating Node on a sidechain. It should be noted that these simplified diagrams do not include node failures, and local time-outs for threshold signing. 

\textit{Subordinate View Process: Coordinating Node: Part 1}, Figure \ref{fig:subordinateviewcoord1}, describes the sequence of events for the Coordinating Node on a Sidechain B to determine whether the core view processing should be undertaken. Walking through the sequence diagram:
\begin{figure*}
 \begin{center}
  \includegraphics[width=\linewidth,height=\textheight,keepaspectratio]{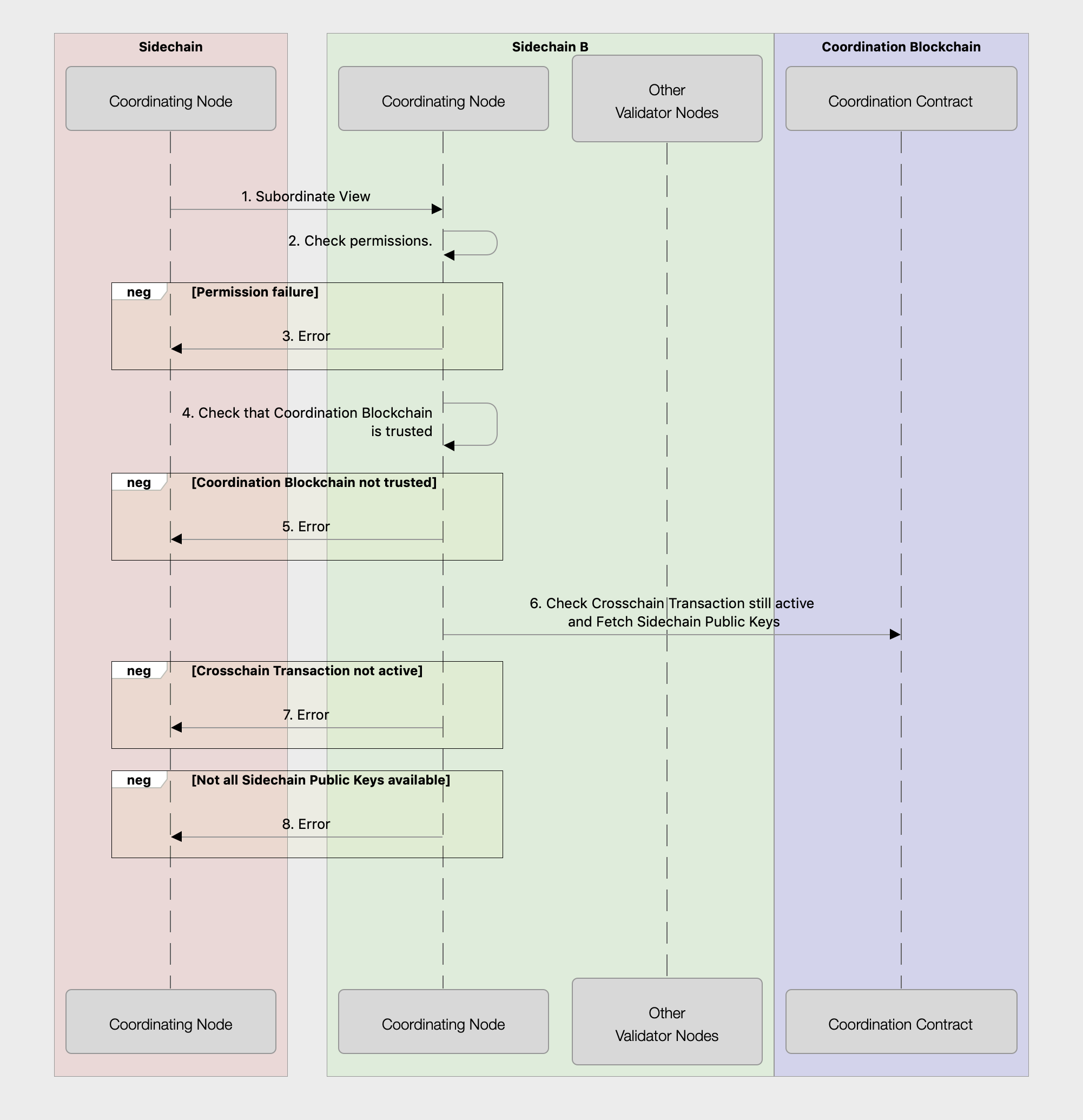}
  \caption{Subordinate View Processing: Coordinating Node Perspective: Part 1}
  \label{fig:subordinateviewcoord1}
 \end{center}
\end{figure*}
\begin{enumerate}
\item The Coordinating Node on a sidechain submits a Subordinate View for processing to the Coordinating Node on Sidechain B. 
\item The node checks whether the account has permission to execute a Subordinate View on this sidechain.
\item An error is returned if the account which signed the Subordinate View is not allowed to execute views on this sidechain.
\item The Coordination Blockchain and Crosschain Coordination Contract address specified in the Subordinate View are checked to see if they are trusted.
\item Return an error if the Coordination Blockchain or the Crosschain Coordination Contract are not trusted by this sidechain.
\item The Coordinating Node on Sidechain B checks that the Crosschain Transaction has been started, has not been committed or ignored, and has not timed-out. While doing this call, the node also fetches Sidechain Public Keys for each Subordinate View called by this Subordinate View's function call from the Crosschain Coordination Contract.
\item An error is returned if the Crosschain Transaction is not still active.
\item An error is returned if all of the Sidechain Public Keys for the sidechains which Subordinate Views are to be submitted to are not available.
\end{enumerate}

\textit{Subordinate View Process: Coordinating Node: Part 2}, Figure \ref{fig:subordinateviewcoord2}, describes the sequence of events for the Coordinating Node on Sidechain B to execute the core view processing. Walking through the sequence diagram:
\begin{figure*}
 \begin{center}
  \includegraphics[width=\linewidth,height=\textheight,keepaspectratio]{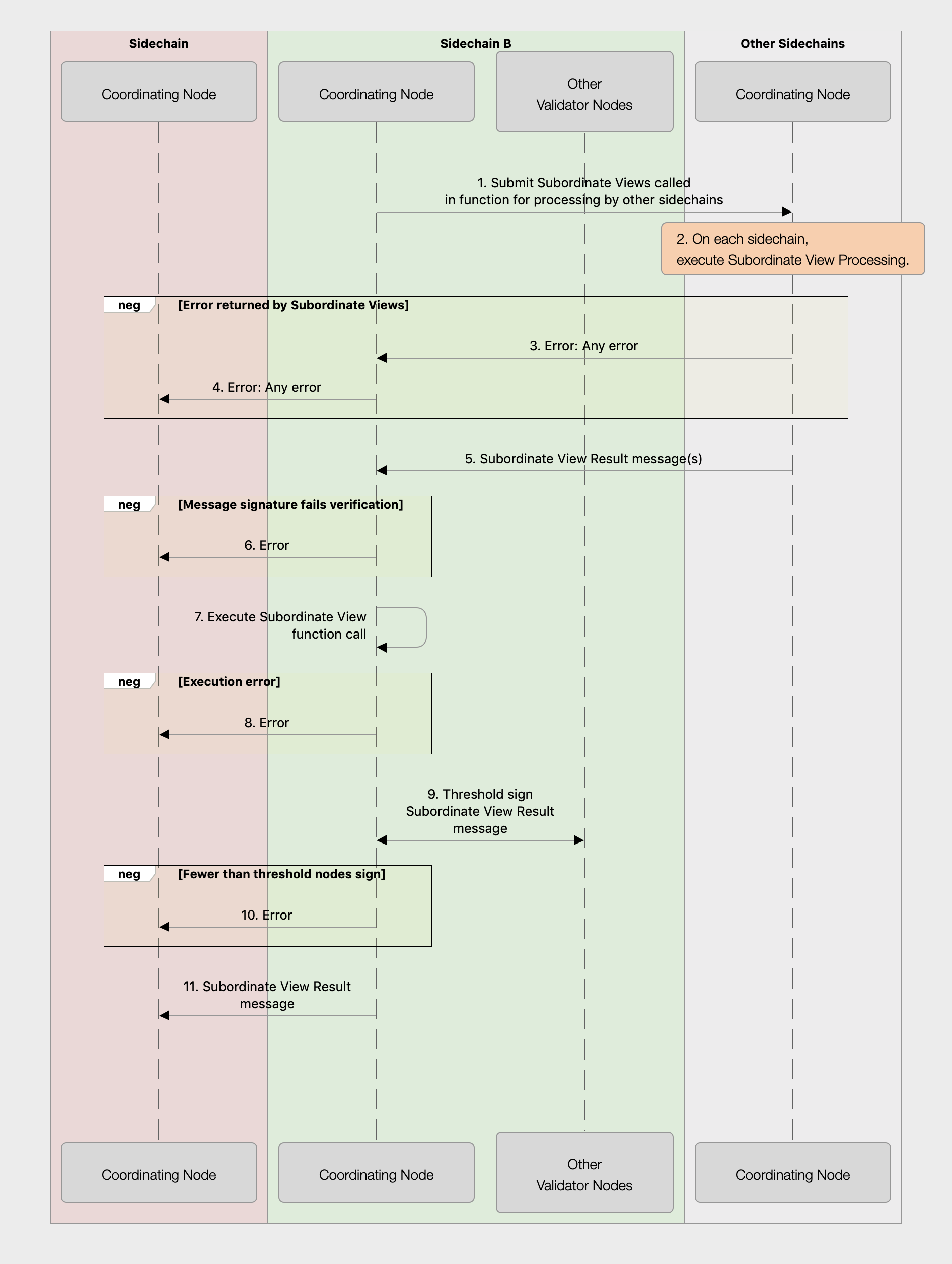}
  \caption{Subordinate View Processing: Coordinating Node Perspective: Part 2}
  \label{fig:subordinateviewcoord2}
 \end{center}
\end{figure*}
\begin{enumerate}
\item The Coordinating Node on Sidechain B submits any Subordinate Views called as a result of the Subordinate View function call being processed to the Coordinating Nodes on the sidechains which the Subordinate Views need to execute.
\item On each sidechain, the Subordinate Views are processed recursively according to the Subordinate View Processing Rules described in this section.
\item An error is returned if any of the Subordinate Views dispatched from this sidechain to other sidechains return an error.
\item Errors from called Subordinate Views are propagated back to the calling sidechain.
\item Assuming no errors are returned by any of the other sidechains and the time-out did not expire, then a Subordinate View Result message will be returned for each Subordinate View submitted to other sidechains. 
\item The signature of each Subordinate View Result message is checked using the Sidechain Public Key of the sidechain the Subordinate View was executed on. An error is returned if the signature on one or more of the Subordinate View Result messages returned from other sidechains fails to verify.
\item The Subordinate View function call to be processed on Sidechain B is executed. When a Subordinate View is called from within the function call, the actual sidechain, contract address and parameter values are compared against the signed values which are the next Subordinate View to be dispatched. The function execution aborts if the values do not match. If they do match, then the return value specified in the Subordinate View Result message is returned to the function.
\item An error is returned if there is an execution error. In addition to the standard Ethereum EVM errors which standard Ethereum contracts can encounter, it is an error if the actual parameters and the signed parameters of a Subordinate View called from the Subordinate View function call being processed do not match.
\item Work with all of the validator nodes on the sidechain to threshold sign a Subordinate View Result message.
\item An error is returned if not enough nodes indicate they are prepared to sign the Subordinate View Result message. In this case, the nodes would have returned error messages indicating they would not sign. Additionally, the nodes may time-out.
\item Send the Subordinate View Result message to the Coordinating Node which submitted the Subordinate View for processing.
\end{enumerate}

\subsubsection{Subordinate View Processing: Other Nodes}
Figure \ref{fig:subordinateviewother1}, shows the sequence diagram for the first half of the processing of a Subordinate View from the perspective of a validator node which is not a Coordinating Node on a sidechain. Walking through the sequence diagram:
\begin{figure*}
 \begin{center}
  \includegraphics[width=\linewidth,height=\textheight,keepaspectratio]{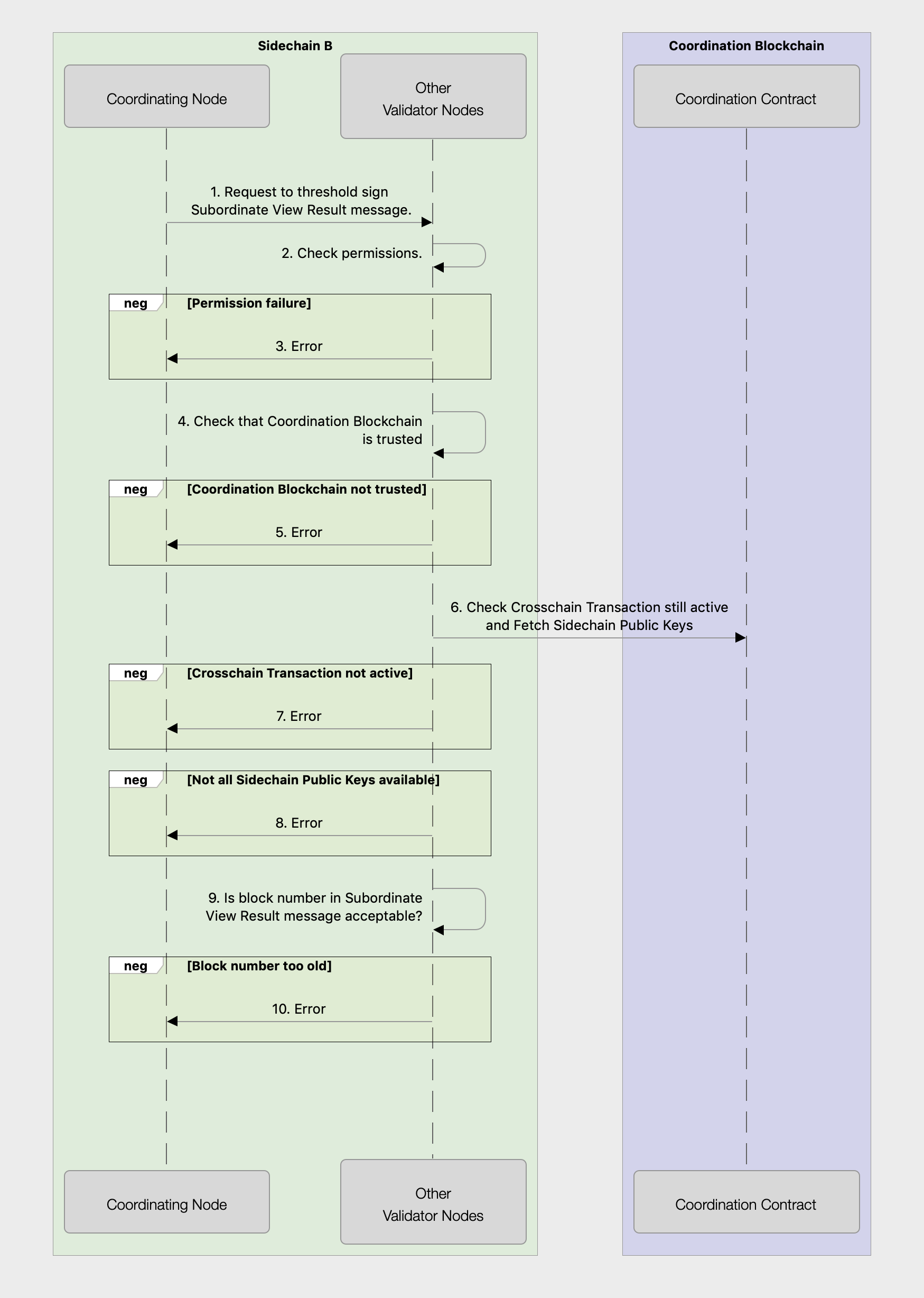}
  \caption{Subordinate View Processing: Nodes Other than the Coordinate Node Perspective: Part 1}
  \label{fig:subordinateviewother1}
 \end{center}
\end{figure*}
\begin{enumerate}
\item The Coordinating Node on the sidechain sends a Subordinate View Result message to be signed to the node. The request includes the Subordinate View which is being processed along with the signed Subordinate View Result messages for all other sidechains which are called as part of the Subordinate View function call.
\item The node checks whether the account has permission to execute a Subordinate View on this sidechain.
\item An error is returned if the account which signed the Subordinate View is not allowed to execute views on this sidechain.
\item The Coordination Blockchain and Crosschain Coordination Contract address specified in the Subordinate View are checked to see if they are trusted.
\item Return an error if the Coordination Blockchain or the Crosschain Coordination Contract are not trusted by this sidechain.
\item The node checks that the Crosschain Transaction has been started, has not been committed or ignored, and has not timed-out. During the same call, the node fetches Sidechain Public Keys for each Subordinate View called by this Subordinate View's function call from the Crosschain Coordination Contract.
\item An error is returned if the Crosschain Transaction is not still active.
\item An error is returned if all of the Sidechain Public Keys for the sidechains for all Subordinate Views called by the function are not available.
\item The node checks the block number specified in the Subordinate View Result message is valid. That is, that the block number is not in the future and is not too old.
\end{enumerate}

\textit{Subordinate View Processing: Nodes Other than the Coordinating Node Perspective: Part 2}, Figure \ref{fig:subordinateviewother2}, shows the sequence diagram for the second half of the processing of a Subordinate View from the perspective of a Validator Node which is not a Coordinating Node on a sidechain. Walking through the sequence diagram:
\begin{figure*}
 \begin{center}
  \includegraphics[width=\linewidth,height=\textheight,keepaspectratio]{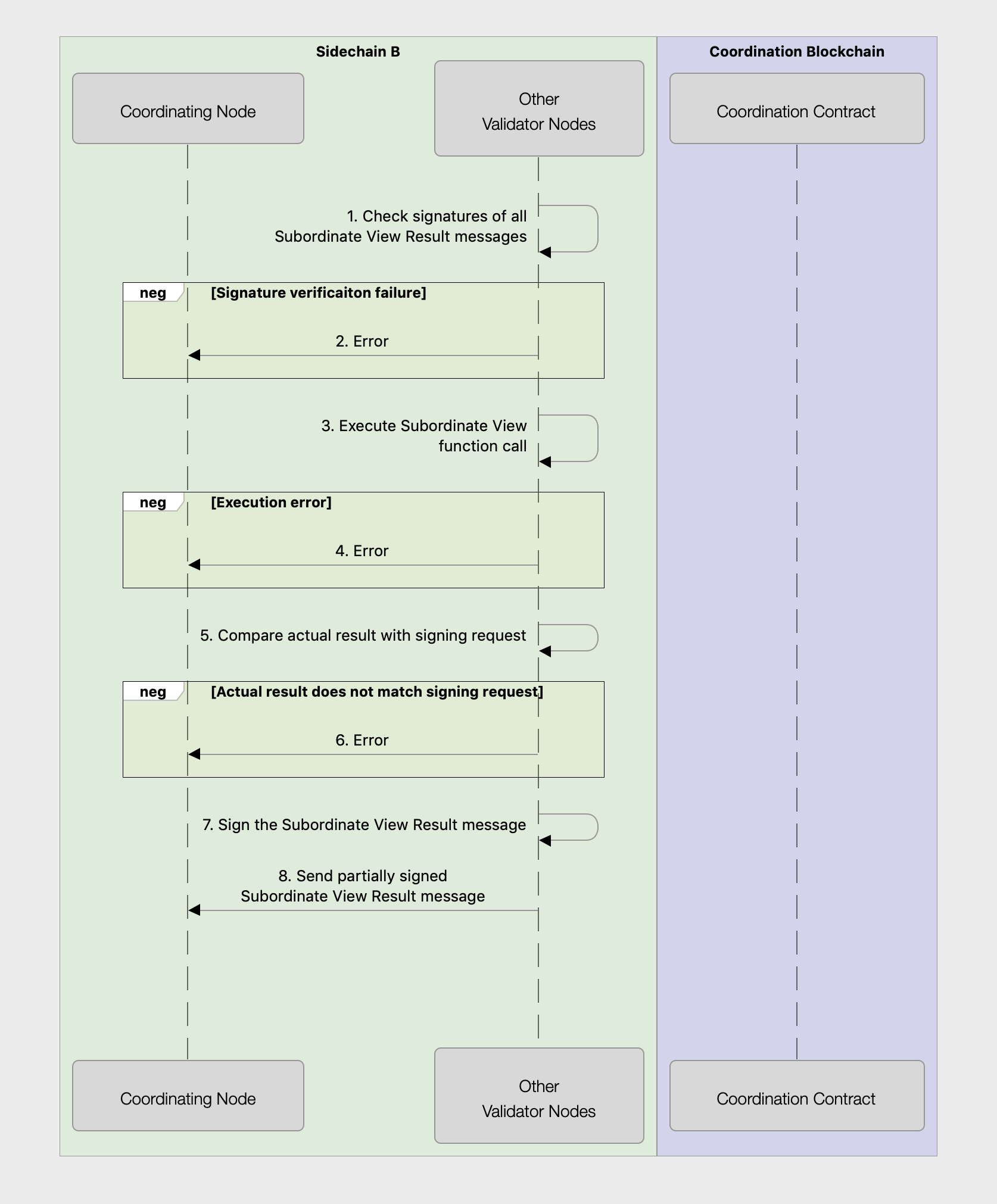}
  \caption{Subordinate View Processing: Nodes Other than the Coordinate Node Perspective: Part 2}
  \label{fig:subordinateviewother2}
 \end{center}
\end{figure*}
\begin{enumerate}
\item The signature of each Subordinate View Result message is checked using the Sidechain Public Key of the sidechain the Subordinate View was executed on. 
\item An error is returned if one or more of the signatures does not verify.
\item The Subordinate View function call to be processed on Sidechain B is executed. When a Subordinate View is called from within the function call, the actual sidechain, contract address and parameter values are compared against the signed values which are the next Subordinate View to be dispatched. The function execution aborts if the values do not match. If they do match, then the return value specified in the Subordinate View Result message is returned to the function.
\item An error is returned if there is an execution error. In addition to the standard Ethereum EVM errors which standard Ethereum contracts can encounter, it is an error if the actual parameters and the signed parameters of a Subordinate View called from the Subordinate View function call being processed do not match.
\item Check that the result shown in the request to sign the Subordinate View Result message matches the result of the Subordinate View function call execution.
\item Return an error if the actual result does not match what the Coordinating Node is requesting be signed.
\item Threshold sign the Subordinate View Result message.
\item Return the partially signed message to the Coordinating Node.
\end{enumerate}

\subsection{Subordinate Transaction Processing}
\label{section:subordinatetransactionprocessing}
Subordinate Transactions can be submitted to a sidechain as a result of the Originating Transaction or other Subordinate Transactions. This section presents how Subordinate Transactions should be processed.

\subsubsection{Subordinate Transaction Processing: Coordinating Node}
Together Figures \ref{fig:subordinatetranscoord1} and \ref{fig:subordinatetranscoord2} show sequence diagrams for the processing a Subordinate Transaction from the perspective of a Coordinating Node on a sidechain. It should be noted that these simplified diagrams do not account for node failures, requests to abort the Crosschain Transaction part way through the sequence, and do not include local time-outs set whilst waiting for the threshold signing process. 

\textit{Subordinate Transaction Processing: Coordinating Node Perspective: Part 1}, Figure \ref{fig:subordinatetranscoord1}, shows the sequence diagram for the first half of the processing of a Subordinate Transaction from the perspective of a Coordinating Node on a sidechain. Walking through the sequence diagram:
\begin{figure*}
 \begin{center}
  \includegraphics[width=\linewidth,height=\textheight,keepaspectratio]{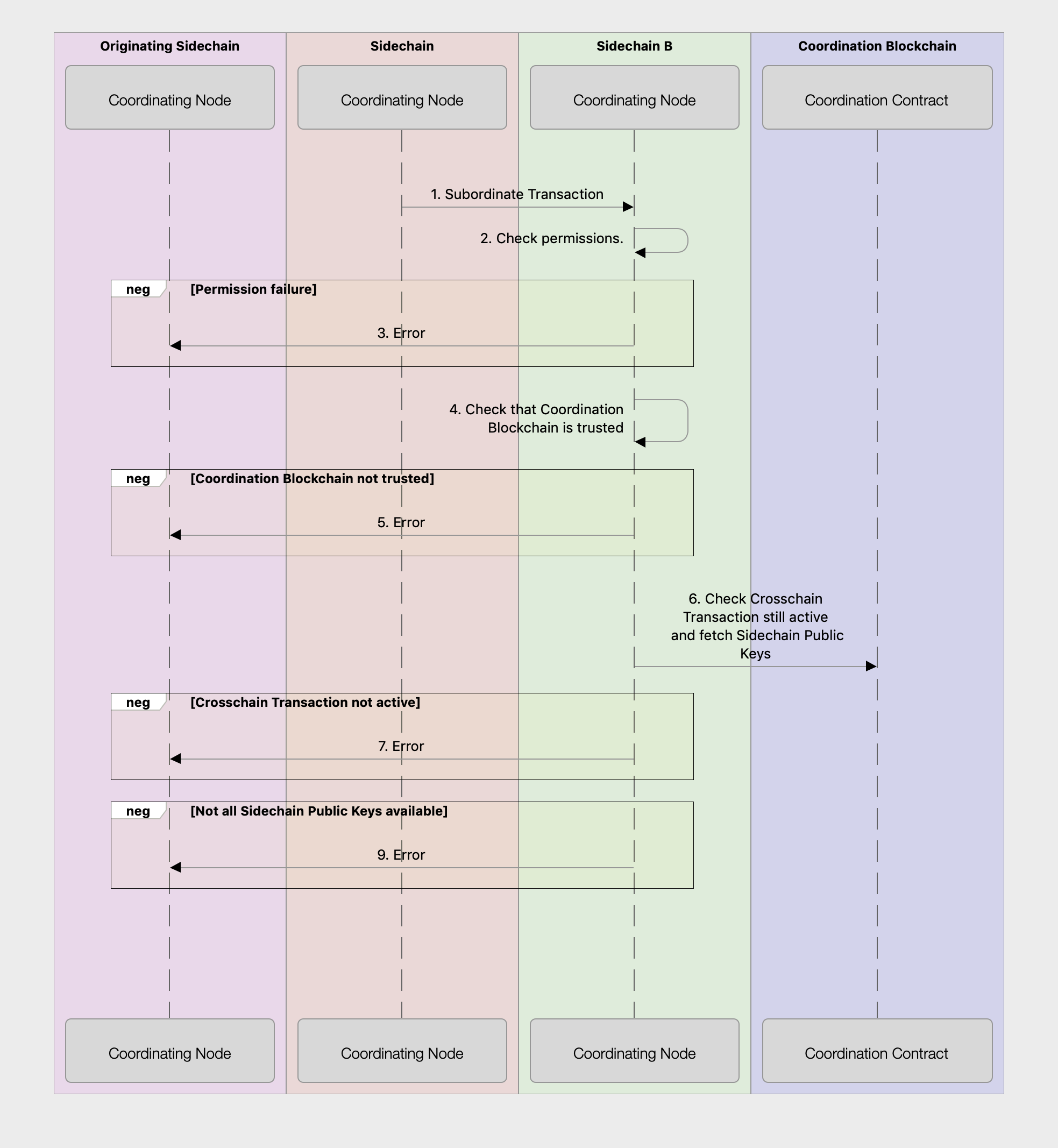}
  \caption{Subordinate Transaction Processing: Coordinating Node Perspective: Part 1}
  \label{fig:subordinatetranscoord1}
 \end{center}
\end{figure*}
\begin{enumerate}
\item The Coordinating Node on a sidechain submits a Subordinate Transaction for processing to the Coordinating Node on Sidechain B. 
\item The Coordinating Node on Sidechain B checks whether the account which signed the transaction has permission to execute transactions on this sidechain.
\item An error is returned to the Coordinating Node on the Originating Sidechain if the account which signed this Subordinate Transaction is not allowed to submit transactions to this sidechain.
\item The Coordination Blockchain and Crosschain Coordination Contract specified in the Subordinate Transaction are checked to see if they are trusted.
\item Return an error to the Coordinating Node on the Originating Sidechain if the Coordination Blockchain or the Crosschain Coordination Contract are not trusted by this sidechain.
\item The Coordinating Node on Sidechain B checks that the Crosschain Transaction has been started, has not been committed or ignored, and has not timed-out. The node also fetches Sidechain Public Keys for each Subordinate View called by this Subordinate View's function call from the Crosschain Coordination Contract.
\item An error is returned to the Coordinating Node on the Originating Sidechain if the Crosschain Transaction is not still active.
\item An error is returned to the Coordinating Node on the Originating Sidechain if all of the Sidechain Public Keys for the sidechains which Subordinate Views are to be called as a result of the Subordinate Transaction function call are not available.
\end{enumerate}

\textit{Subordinate Transaction Processing: Coordinating Node Perspective: Part 2}, Figure \ref{fig:subordinatetranscoord2}, shows the sequence diagram for the second half of the processing of a Subordinate Transaction from the perspective of a Coordinating Node on a sidechain. Walking through the sequence diagram:
\begin{figure*}
 \begin{center}
  \includegraphics[width=\linewidth,height=\textheight,keepaspectratio]{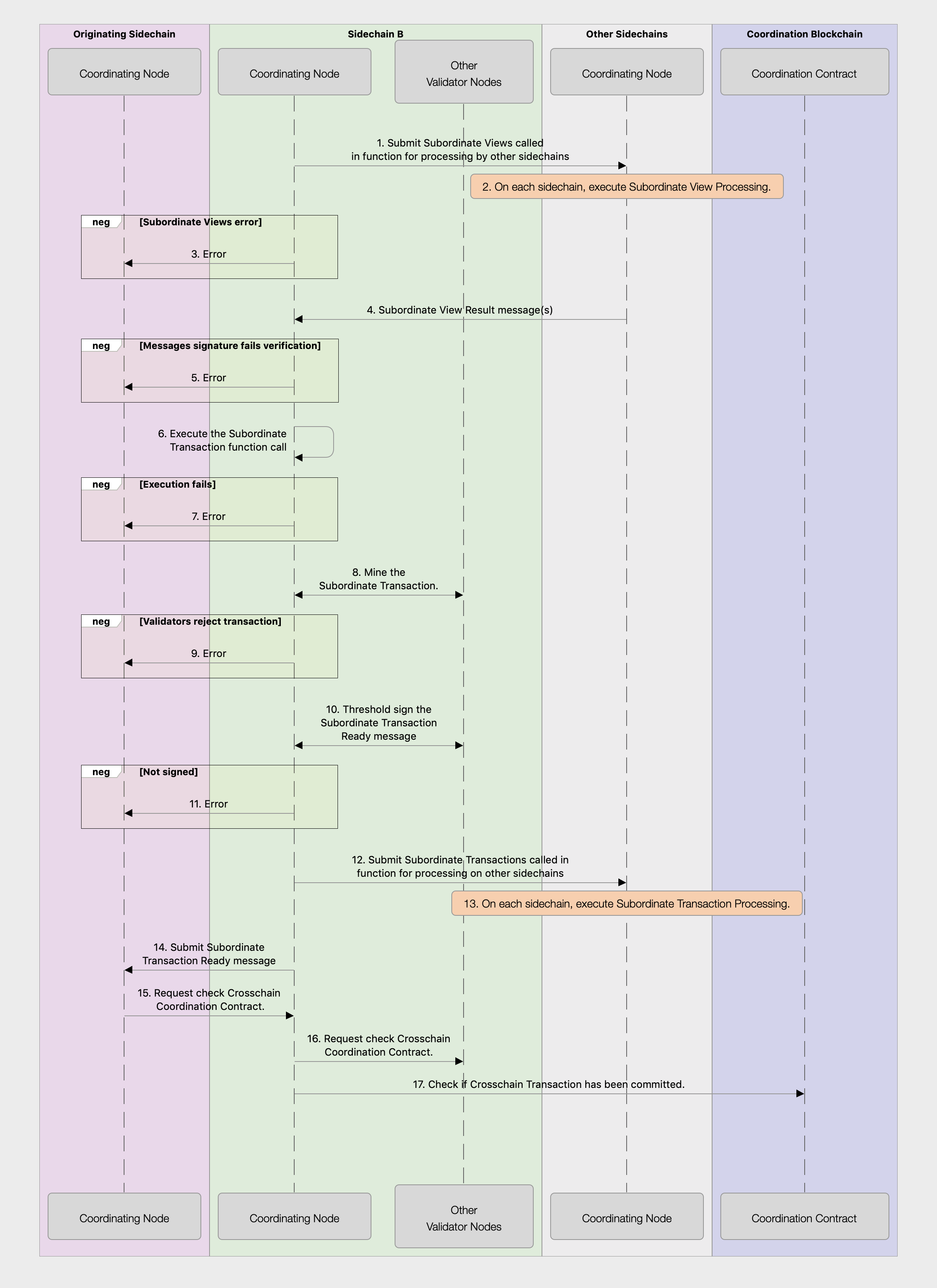}
  \caption{Subordinate Transaction Processing: Coordinating Node Perspective: Part 2}
  \label{fig:subordinatetranscoord2}
 \end{center}
\end{figure*}
\begin{enumerate}
\item The Coordinating Node on Sidechain B submits any Subordinate Views called as a result of the Subordinate Transaction function call it is processing to the Coordinating Nodes on the sidechains which the Subordinate Views need to execute.
\item On each sidechain, the Subordinate Views are processed recursively according to the Subordinate View Processing Rules described in Section  \ref{section:subordinateviewprocessing}. 
\item An error is returned to the Coordinating Node on the Originating Sidechain if any of the Subordinate Views returns an error.
\item The Coordination Nodes on the sidechains which have executed the Subordinate Views returns a threshold signed Subordinate View Result message.
\item An error is returned to the Coordinating Node on the Originating Sidechain if a Subordinate View Result message for each dispatched Subordinate View is not returned. The transaction fails if the signatures on all of the Subordinate Views can not be verified.
\item Execute the function call in the Subordinate Transaction. When a Subordinate View or Subordinate Transaction is called from within the function call, the actual sidechain, contract address and parameter values are compared against the signed values which are the next Subordinate Transaction or View to be dispatched. The function execution aborts if the values do not match. If they do match, then for the Subordinate Views, the return value specified in the Subordinate View Result message is returned to the function.
\item An error is returned to the Coordinating Node on the Originating Sidechain if the function fails to execute to completion.
\item Distribute the Subordinate Transaction and Subordinate View Result messages to all validators and have the transaction mined according to the sidechain's consensus algorithm. The contract is locked when the transaction is included in the blockchain.
\item An error is returned to the Coordinating Node on the Originating Sidechain if the validators reject the transaction.
\item The Coordinating Node works with all of the validator nodes on the sidechain to threshold sign a Subordinate Transaction Ready message.
\item An error is returned if not enough nodes indicate they are prepared to sign the Subordinate Transaction Ready message. The nodes will respond with error messages indicating why they do not want to sign.
\item Any Subordinate Transactions called from the the Originating Transaction are dispatched to the appropriate sidechain.
\item On each sidechain, Subordinate Transactions are processed recursively according to the Subordinate Transaction Processing rules described in Section \ref{section:subordinatetransactionprocessing}.
\item The Subordinate Transaction Ready message is sent from the Coordinating Node on Sidechain B to the Coordinating Node on the Originating Sidechain.
\item Once the Crosschain Transaction Commit message has been verified and accepted by the Crosschain Coordination Contract, the Crosschain Transaction is ready to be committed on all sidechains. The Coordinating Node on the Originating Sidechain sends a message requesting that all nodes check the Crosschain Coordination Contract.
\item The Coordinating Node on Sidechain B trusts the Coordinating Node on the Originating Sidechain because they are a part of the same Multichain Node. As such, it forwards the message requesting that all nodes check the Crosschain Coordination Contract immediately.
\item The Coordinating Node on Sidechain B checks the Crosschain Coordination Contract to see if the Crosschain Transaction was committed or ignored. The Coordinating Node applies the state updates if the transaction was committed. If the state is \texttt{Committed} or \texttt{Ignored} the node unlocks the contract.
\end{enumerate}

\subsubsection{Subordinate Transaction Processing: Other Nodes}
Together Figures \ref{fig:subordinatetransother1} and \ref{fig:subordinatetransother2} show sequence diagrams for the processing a Subordinate Transaction from the perspective of a node which is not the Coordinating Node on a sidechain. It should be noted that these simplified diagrams do not account for node failures. 

\textit{Subordinate Transaction Processing: Other Node Perspective: Part 1}, Figure \ref{fig:subordinatetransother1}, shows the sequence diagram for first half of the processing of a Subordinate Transaction from the perspective of a node other than the Coordinating Node on a sidechain. Walking through the sequence diagram:
\begin{figure*}
 \begin{center}
  \includegraphics[width=\linewidth,height=\textheight,keepaspectratio]{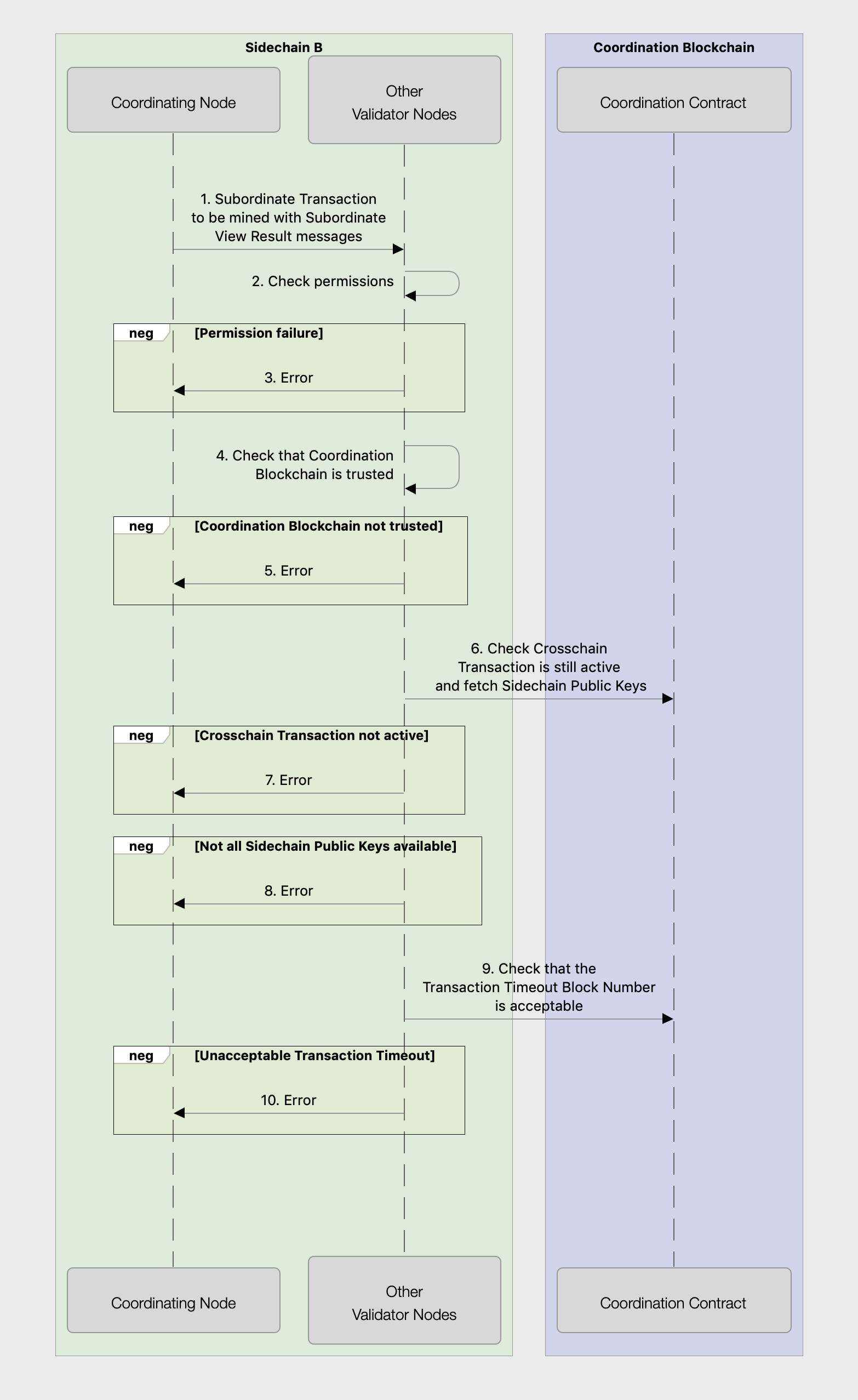}
  \caption{Subordinate Transaction Processing: Other Node Perspective: Part 1}
  \label{fig:subordinatetransother1}
 \end{center}
\end{figure*}
\begin{enumerate}
\item The Coordinating Node on Sidechain B sends a Subordinate Transaction to be mined, along with the associated Subordinate View Result messages to all validator nodes on Sidechain B. 
\item Each validator node checks whether the account which signed the transaction has permission to execute transactions on this sidechain.
\item An error is returned to the Coordinating Node if the account which signed this Subordinate Transaction is not allowed to submit transactions to this sidechain.
\item The Coordination Blockchain and Crosschain Coordination Contract specified in the Subordinate Transaction are checked to see if they are trusted.
\item Return an error to the Coordinating Node if the Crosschain Coordination Contract or the Coordination Blockchain are not trusted by this sidechain.
\item Each validator node checks that the Crosschain Transaction has been started, has not been committed or ignored, and has not timed-out. The node fetches Sidechain Public Keys for each Subordinate View called by this Subordinate Transaction's function call from the Crosschain Coordination Contract.
\item An error is returned to the Coordinating Node if the Crosschain Transaction is not still active.
\item An error is returned to the Coordinating Node if all of the Sidechain Public Keys for the sidechains which Subordinate Views are to be called as a result of the Subordinate Transaction function call are not available.
\item Each validator node checks that the Transaction Timeout Block Number, the global time-out, is an acceptable value. That is, it checks that the block number is equal to or less than the amount of time it is prepared to have a contract locked for. 
\item The validator returns an error if it finds the Transaction Timeout Block Number to be unacceptable.
\end{enumerate}

\textit{Subordinate Transaction Processing: Other Node Perspective: Part 2}, Figure \ref{fig:subordinatetransother2}, shows the sequence diagram for the second half of the processing of a Subordinate Transaction from the perspective of a node other than the Coordinating Node on a sidechain. Walking through the sequence diagram:
\begin{figure*}
 \begin{center}
  \includegraphics[width=\linewidth,height=\textheight,keepaspectratio]{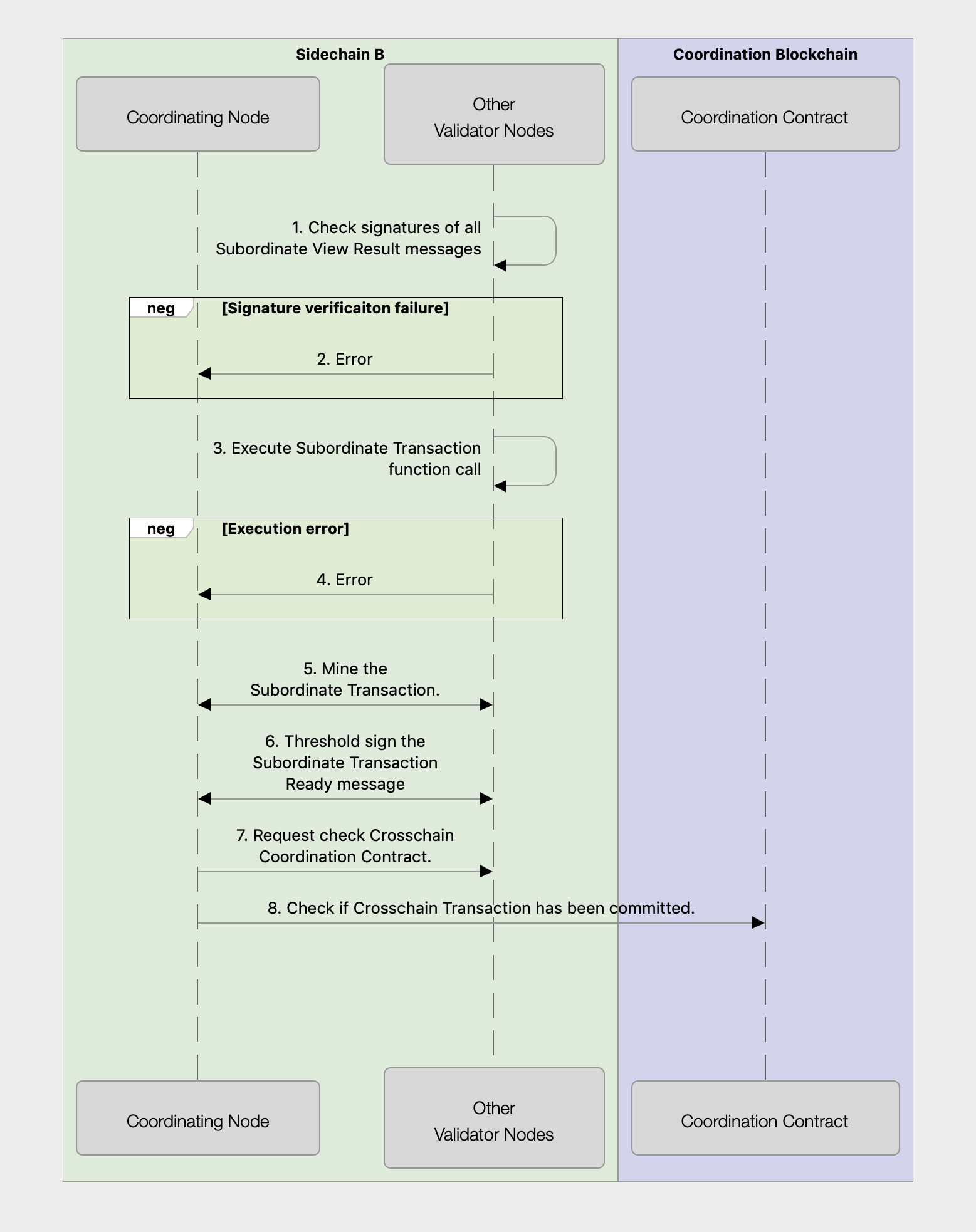}
  \caption{Subordinate Transaction Processing: Other Node Perspective: Part 2}
  \label{fig:subordinatetransother2}
 \end{center}
\end{figure*}
\begin{enumerate}
\item Each validator node checks the signature of each Subordinate View Result message.
\item An error is returned to the Coordinating Node if any of the signatures can not be verified.
\item Execute the function call in the Subordinate Transaction. When a Subordinate View or Subordinate Transaction is called from within the function call, the actual sidechain, contract address and parameter values are compared against the signed values which are the next Subordinate Transaction or View to be dispatched. The function execution aborts if the values do not match. If they do match, then for the Subordinate Views, the return value specified in the Subordinate View Result message is returned to the function.
\item An error is returned to the Coordinating Node if the function fails to execute to completion.
\item The consensus algorithm specific mining algorithm completes.
\item The Coordinating Node sends the Subordinate Transaction Ready message to be signed. The message contains the Originating Sidechain Identifier, the Crosschain Transaction Identifier, the Coordination Blockchain Identifier, the Crosschain Coordination Contract address, the Sidechain Identifier of the sidechain that the Subordinate Transaction has to be executed on and the hash of the Subordinate Transaction. Each validator threshold signs the message and returns it to the Coordinating Node if a transaction which matches the message has been finalised.
\item Once the Crosschain Transaction Commit message has been verified and accepted by the Crosschain Coordination Contract, the Crosschain Transaction is ready to be committed on all sidechains. The Coordinating Node on the Originating Sidechain sends a message requesting that all nodes check the Crosschain Coordination Contract. This message is forwarded from the Coordinating Node on Sidechain B to all validators on the sidechain.
\item Each validator checks the Crosschain Coordination Contract to see if the Crosschain Transaction was committed or ignored. The Coordinating Node applies the state updates if the transaction was committed. If the state is \texttt{Committed} or \texttt{Ignored} the node unlocks the contract.
\end{enumerate}

\subsection{Crosschain (Originating) Transaction Processing}
Applications submit Originating Transactions to sidechains. The sidechains are designated Originating Sidechains for the purpose of the Crosschain Transaction. The processing of the Originating Transaction, and all Subordinate Transactions and Views within the Originating Transaction constitutes processing of the entire Crosschain Transaction. This section presents how Originating Transactions should be processed.

\subsubsection{Crosschain (Originating) Transaction Processing: Coordinating Node}
Together Figures \ref{fig:originatingtranscoord1}, \ref{fig:originatingtranscoord2} and \ref{fig:originatingtranscoord3} show sequence diagrams for the processing an Originating Transaction from the perspective of a Coordinating Node on the Originating Sidechain. It should be noted that these simplified diagrams do not account for node failures, requests to abort the Crosschain Transaction part way through the sequence, and do not include local time-outs set whilst waiting for the threshold signing process. In the diagrams, while the transaction is active, any action which results in Crosschain Transaction failed could then trigger the creation of a Crosschain Transaction Ignore message between the nodes on the Originating Sidechain and submission of that message to the Crosschain Coordination Contract.

\textit{Originating Transaction Process: Coordinating Node: Part 1}, Figure \ref{fig:originatingtranscoord1}, describes the sequence of events for the Coordinating Node on the Originating Sidechain to determine if the Crosschain Transaction should be started. Walking through the sequence diagram:
\begin{figure*}
 \begin{center}
  \includegraphics[width=\linewidth,height=\textheight,keepaspectratio]{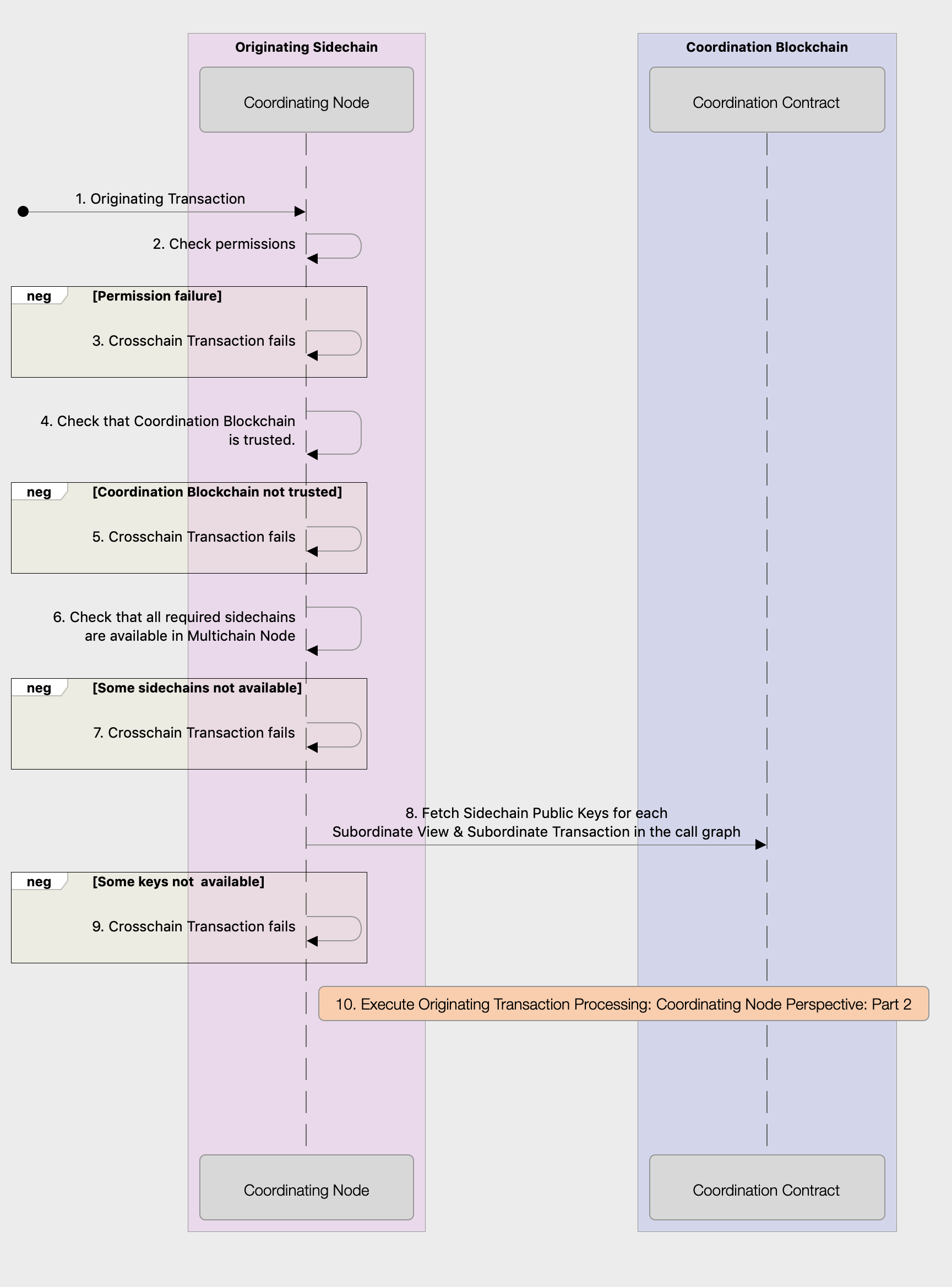}
  \caption{Originating Transaction Processing: Coordinating Node Perspective: Part 1}
  \label{fig:originatingtranscoord1}
 \end{center}
\end{figure*}
\begin{enumerate}
\item The application submits the Originating Transaction to the Coordinating Node on the Originating Sidechain.
\item The Coordinating Node on the Originating Sidechain checks whether the account which signed the transaction has permission to execute transactions on this sidechain.
\item The transaction fails if the account which signed this Originating Transaction is not allowed to submit transactions to this sidechain.
\item The Coordinating Node on the Originating Sidechain checks that the sidechain or blockchain specified by the Coordination Blockchain Identifier in the Originating Transaction is available in Multichain Node and trusted to be a Coordination Blockchain. The Crosschain Coordination Contract address is checked to ensure it is trusted.
\item The transaction fails if the Coordination Blockchain is not available or not trusted or if the Crosschain Coordination Contract is not trusted.
\item The Coordinating Node on the Originating Sidechain checks that the Multichain Node on which it is running contains validator nodes on all of the sidechains represented by the Subordinate View and Subordinate Transaction in the entire Subordinate View and Transaction tree. The Multichain Node does this to ensure the entire Crosschain Transaction will be able to execute.
\item The transaction fails if some of the required sidechains are not available in the Multichain Node.
\item The Coordinating Node on the Originating Sidechain fetches Sidechain Public Keys for each Subordinate View and Subordinate Transaction in the entire Subordinate View and Transaction tree. This is done to ensure all public keys are available.
\item The transaction fails if some of Sidechain Public Keys are not available.
\end{enumerate}

\textit{Originating Transaction Process: Coordinating Node: Part 2}, Figure \ref{fig:originatingtranscoord2}, describes the sequence of events in which the Coordinating Node on the Originating Sidechain starts and then later commits the Crosschain Transaction. Walking through the sequence diagram:
\begin{figure*}
 \begin{center}
  \includegraphics[width=\linewidth,height=\textheight,keepaspectratio]{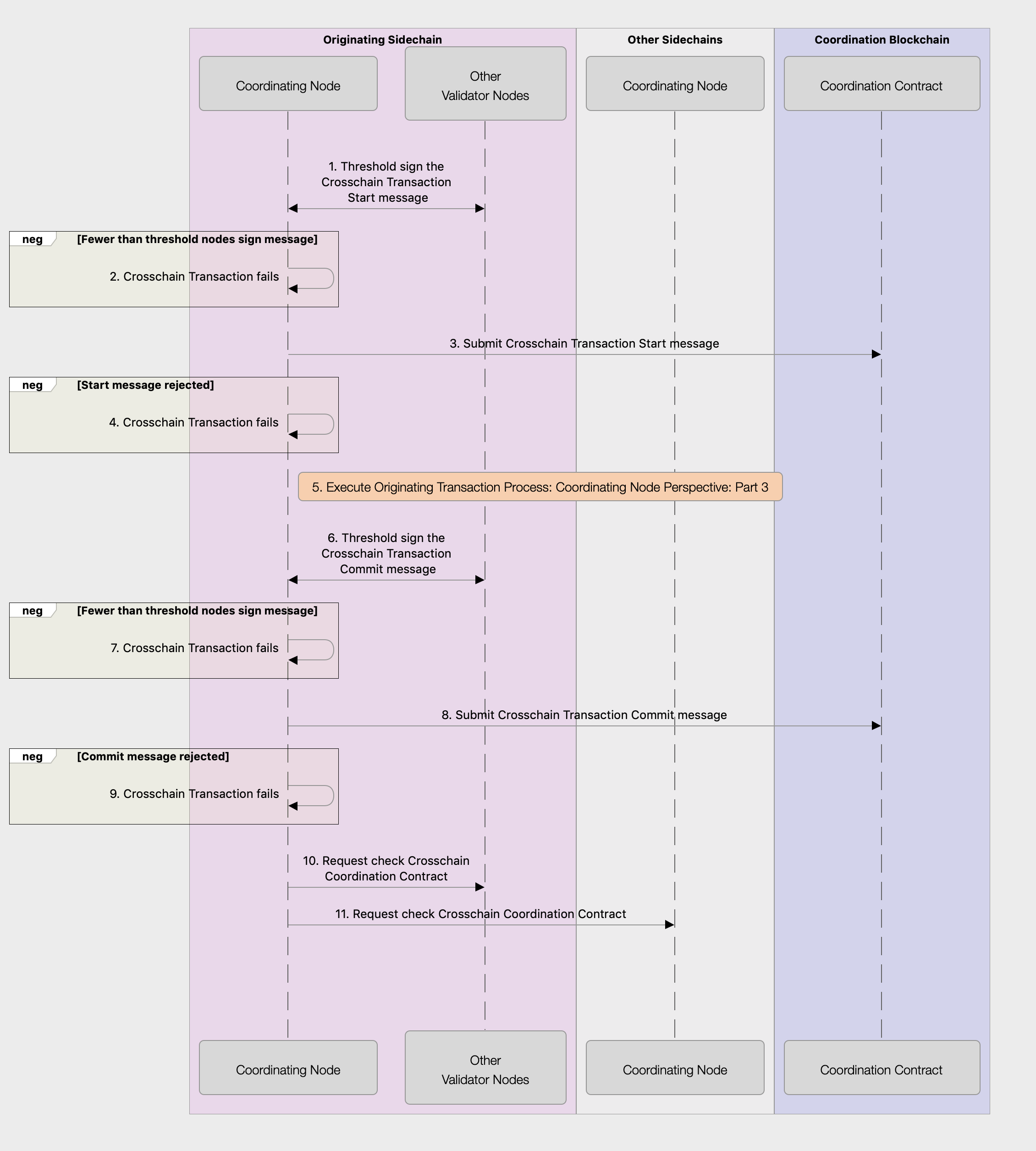}
  \caption{Originating Transaction Processing: Coordinating Node Perspective: Part 2}
  \label{fig:originatingtranscoord2}
 \end{center}
\end{figure*}
\begin{enumerate}
\item The Coordinating Node on the Originating Sidechain works with other validators on the Originating Sidechain to threshold sign the Crosschain Transaction Start message.
\item The Crosschain Transaction fails if fewer than the threshold number of nodes sign the Crosschain Transaction Start message prior to a local time-out expiring.
\item The Crosschain Transaction Start message is submitted to the Crosschain Coordination Contract. The message contains the Originating Sidechain Identifier, Crosschain Transaction Identifier and the Crosschain Transaction Timeout. Assuming the message is accepted, an entry for the Crosschain Transaction is created. The message is rejected if the combination of Crosschain Transaction Identifier and Originating Sidechain already exist, or if the threshold signature can not be verified by the Sidechain Public Key, or if the Crosschain Transaction Timeout is greater than the largest time-out the contract is configured to allow.
\item The Crosschain Transaction fails if the Crosschain Coordination Contract rejects the Crosschain Transaction Start message.
\item The core transaction processing then occurs. See Figure \ref{fig:originatingtranscoord3}. At the end of this process the Originating Transaction and all Subordinate Transactions are ready to be committed.
\item The Coordinating Node on the Originating Sidechain works with other validators on the Originating Sidechain to threshold sign the Crosschain Transaction Commit message.
\item The Crosschain Transaction fails if fewer than the threshold number of nodes sign the Crosschain Transaction Commit message prior to a local time-out expiring.
\item The Crosschain Transaction Commit message is submitted to the Crosschain Coordination Contract. Assuming the message is accepted, the entry for the Crosschain Transaction is updated to indicate it is committed. The message is rejected if the combination of Crosschain Transaction Identifier and Originating Sidechain do not exist, or if the threshold signature can not be verified by the Sidechain Public Key, if the transaction has already been marked as \texttt{Committed} or \texttt{Ignored}, or if the block number is greater than the Transaction Timeout Block Number.
\item The Crosschain Transaction fails if the Crosschain Coordination Contract rejects the Crosschain Transaction Commit message.
\item The Coordinating Node on the Originating Sidechain requests other nodes on the Originating Sidechain check the Crosschain Coordination Contract to see that the transaction has been committed.
\item The Coordinating Node on the Originating Sidechain requests that Coordinating Nodes on other sidechain check the Crosschain Coordination Contract to see that the transaction has been committed.
\end{enumerate}

\textit{Originating Transaction Process: Coordinating Node: Part 3}, Figure \ref{fig:originatingtranscoord3}, describes the sequence of events in which the Coordinating Node on the Originating Sidechain requests the results of Subordinate Views, executes the Originating Transaction function call, and submits any Subordinate Transactions which result from the function call. Walking through the sequence diagram:
\begin{figure*}
 \begin{center}
  \includegraphics[width=\linewidth,height=\textheight,keepaspectratio]{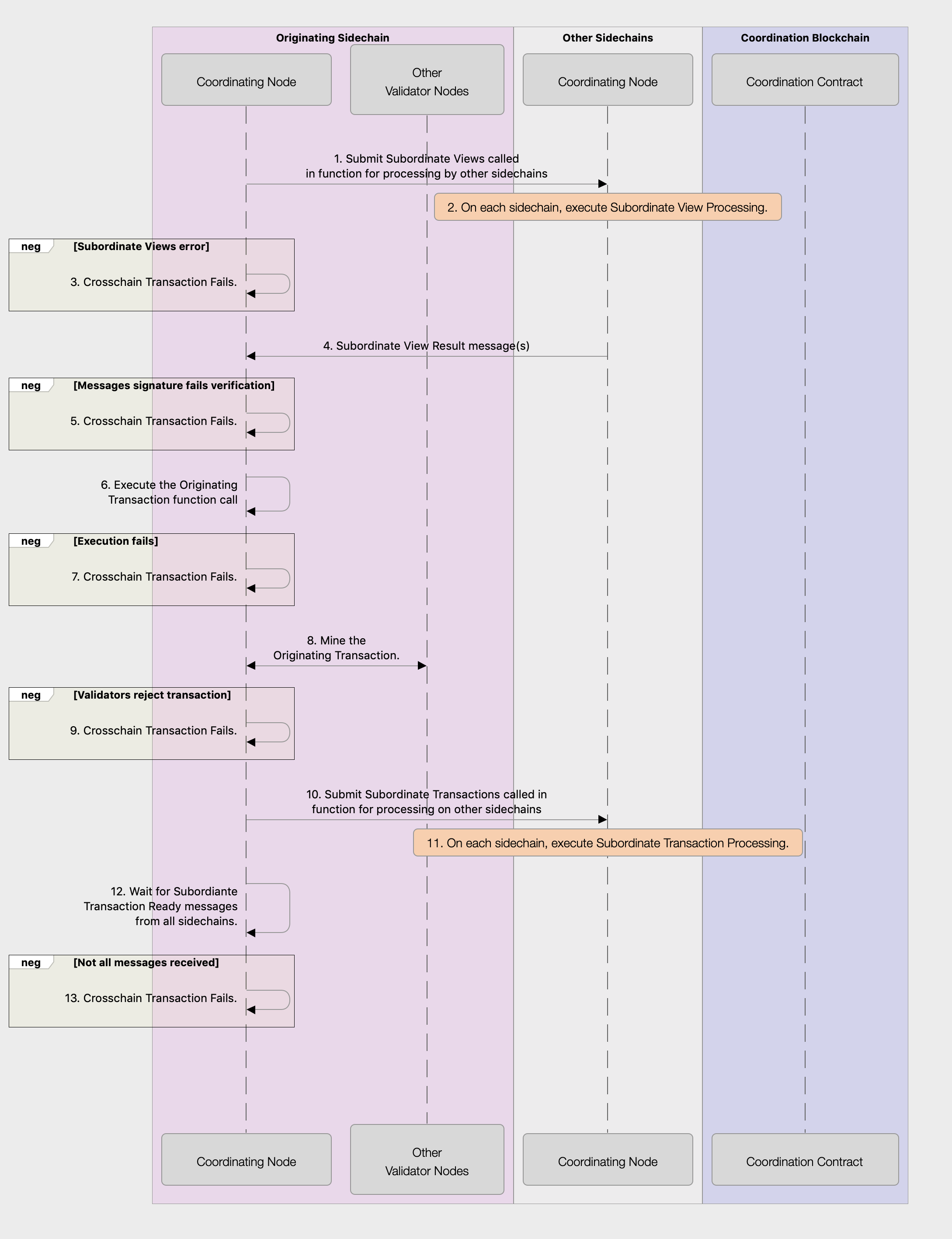}
  \caption{Originating Transaction Processing: Coordinating Node Perspective: Part 3}
  \label{fig:originatingtranscoord3}
 \end{center}
\end{figure*}
\begin{enumerate}
\item The Coordination Node on the Originating Sidechain dispatches all of the Subordinate Views which are listed in the Subordinate Transaction and View hierarchy immediately below the Originating Transaction to the sidechains they should be executed on. 
\item Each sidechain processes the Subordinate View according to the Subordinate View Processing rules described in Section \ref{section:subordinateviewprocessing}. 
\item The Crosschain Transaction Fails if any of the Subordinate Views returns an error.
\item The Coordination Nodes on the sidechains which have executed the Subordinate Views returns a threshold signed Subordinate View Result message.
\item The Crosschain Transaction Fails if a Subordinate View Result message for each dispatched Subordinate View is not returned. The transaction fails if the signatures on all of the Subordinate Views can not be verified.
\item Execute the function call in the Originating Transaction. When a Subordinate View or Subordinate Transaction is called from within the function call, the actual sidechain, contract address and parameter values are compared against the signed values which are the next Subordinate Transaction or View to be dispatched. The function execution aborts if the values do not match. If they do match, then for the Subordinate Views, the return value specified in the Subordinate View Result message is returned to the function.
\item The transaction fails if the function fails to execute to completion.
\item Distribute the Originating Transaction and Subordinate View Result messages to all validators and have the transaction mined according to the sidechain's consensus algorithm.
\item The Crosschain Transaction fails if the validators reject the transaction.
\item Any Subordinate Transactions called from the the Originating Transaction are dispatched to the appropriate sidechain.
\item On each sidechain, Subordinate Transactions are processed recursively according to the Subordinate Transaction Processing rules described in Section \ref{section:subordinatetransactionprocessing}.
\item The Coordinating Node on the Originating Sidechain waits a Subordinate Transaction Ready messages is received for each of the dispatched Subordinate Transactions.
\item The Crosschain Transaction fails if not all of the Subordinate Transaction Ready messages are received prior to a local time-out, or if the signature on any of the messages fails.
\end{enumerate}

\subsubsection{Crosschain (Originating) Transaction Processing: Other Nodes}
Together Figures \ref{fig:originatingtransother1} and \ref{fig:originatingtransother2} show sequence diagrams for the processing an Originating Transaction from the perspective of a node which is not the Coordinating Node on the Originating Sidechain. It should be noted that these simplified diagrams do not account for node failures. 

\textit{Originating Transaction Processing: Other Node Perspective: Part 1}, Figure \ref{fig:originatingtransother1}, shows the sequence diagram for first half of the processing of an Originating Transaction from the perspective of a node other than the Coordinating Node on the Originating Sidechain. Walking through the sequence diagram:
\begin{figure*}
 \begin{center}
  \includegraphics[width=\linewidth,height=\textheight,keepaspectratio]{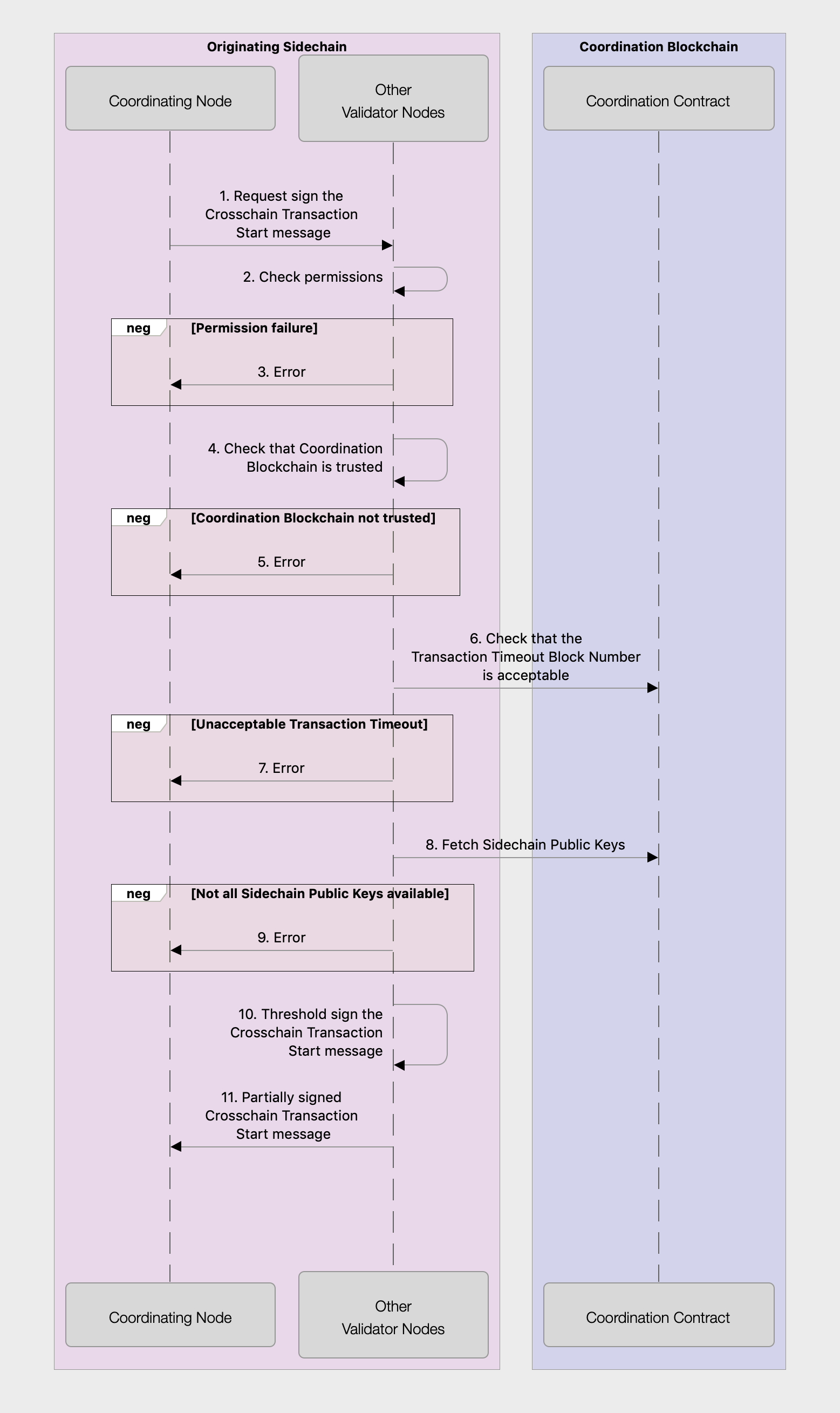}
  \caption{Originating Transaction Processing: Other Node Perspective: Part 1}
  \label{fig:originatingtransother1}
 \end{center}
\end{figure*}
\begin{enumerate}
\item The Coordinating Node sends a request to threshold sign the Crosschain Transaction Start message and the Originating Transaction to all validator nodes on the Originating Sidechain. 
\item Each validator node checks whether the account which signed the transaction has permission to execute transactions on this sidechain.
\item An error is returned to the Coordinating Node if the account which signed this Originating Transaction is not allowed to submit transactions to this sidechain.
\item The Coordination Blockchain and Crosschain Coordination Contract specified in the Originating Transaction are checked to see if they are trusted.
\item Return an error to the Coordinating Node if the Crosschain Coordination Contract or the Coordination Blockchain are not trusted by this sidechain.
\item Each validator node checks that the Crosschain Transaction Time-out proposed in the Crosschain Transaction Start message, the global time-out, is an acceptable value. That is, it checks that the number of blocks is equal to or less than the amount of time it is prepared to have a contract locked for. 
\item The validator returns an error if it finds the Crosschain Transaction Timeout to be unacceptable.
\item Each validator fetches Sidechain Public Keys for all Subordinate Views and Subordinate Transactions in the call graph starting from the Originating Transaction from the Crosschain Coordination Contract.
\item An error is returned to the Coordinating Node if all of the Sidechain Public Keys are not available.
\item Threshold sign the Crosschain Transaction Start message.
\item Return the partially signed Crosschain Transaction Start message to the Coordinating Node.
\end{enumerate}

\textit{Originating Transaction Processing: Other Node Perspective: Part 2}, Figure \ref{fig:originatingtransother2}, shows the sequence diagram for the second half of the processing of an Originating Transaction from the perspective of a node other than the Coordinating Node on the Originating Sidechain. Walking through the sequence diagram:
\begin{figure*}
 \begin{center}
  \includegraphics[width=\linewidth,height=\textheight,keepaspectratio]{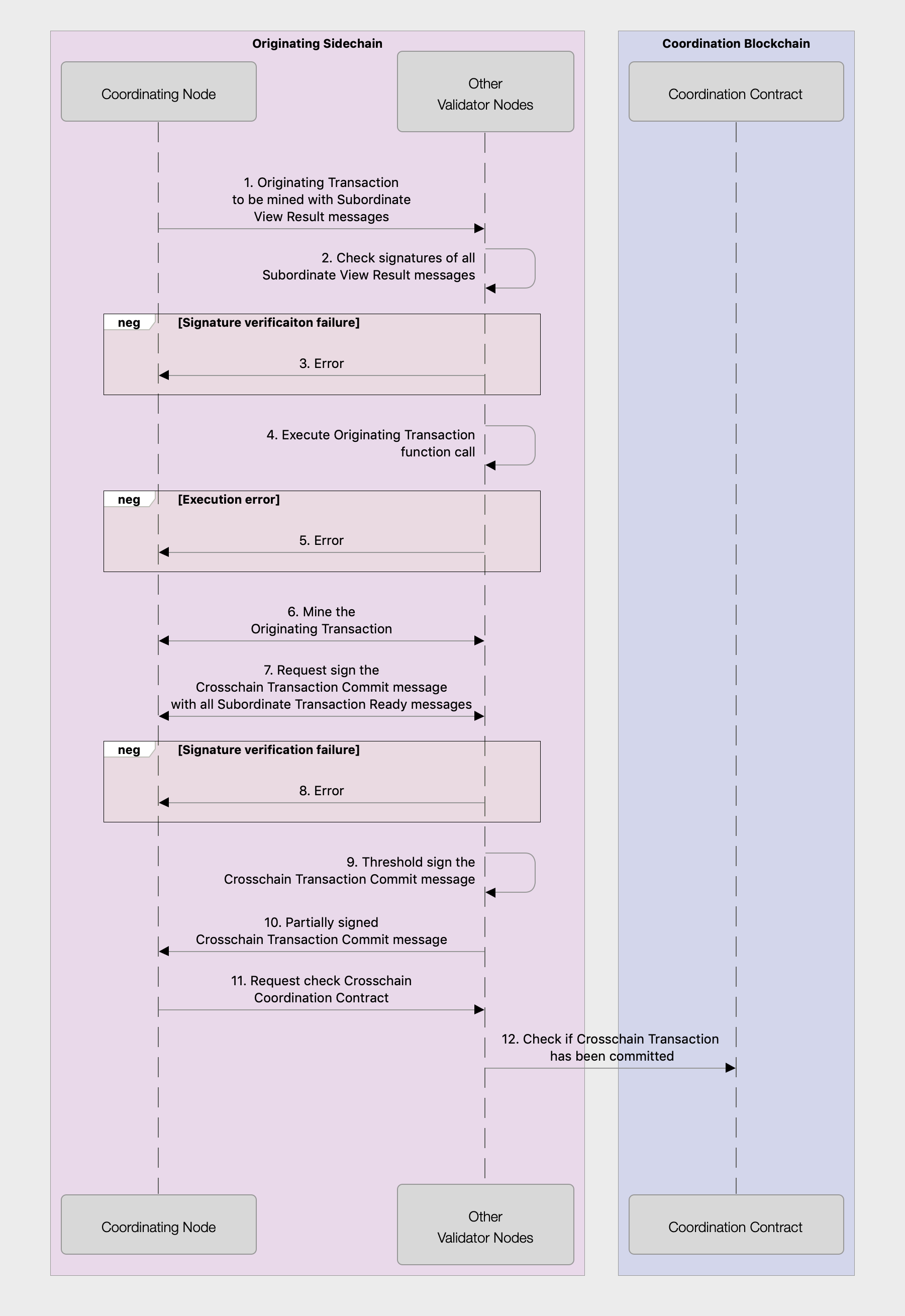}
  \caption{Originating Transaction Processing: Other Node Perspective: Part 2}
  \label{fig:originatingtransother2}
 \end{center}
\end{figure*}
\begin{enumerate}
\item The Coordinating Node sends the Originating Transaction to be mined, along with the associated Subordinate View Result messages to all validator nodes on the Originating Sidechain. 
\item Each validator node checks the signature of each Subordinate View Result message.
\item An error is returned to the Coordinating Node if any of the signatures can not be verified.
\item Execute the function call in the Originating Transaction. When a Subordinate View or Subordinate Transaction is called from within the function call, the actual sidechain, contract address and parameter values are compared against the signed values which are the next Subordinate Transaction or View to be dispatched. The function execution aborts if the values do not match. If they do match, then for the Subordinate Views, the return value specified in the Subordinate View Result message is returned to the function.
\item An error is returned to the Coordinating Node if the function fails to execute to completion.
\item The consensus algorithm specific mining algorithm completes.
\item Once all Subordinate Transactions have returned Subordinate Transaction Ready messages to the Coordinating Node on the Originating Blockchain, it sends the Crosschain Transaction Commit message to be signed. The Subordinate Transaction Ready messages for all Subordinate Transactions are attached to the request. 
\item The signatures on each Subordinate Transaction Ready message is checked. An error is returned if any of the signatures fail to verify.
\item The validator threshold signs the Crosschain Transaction Commit message.
\item The partially signed Crosschain Transaction Commit message is returned to the Coordinating Node. 
\item Once the Crosschain Transaction Commit message has been verified and accepted by the Crosschain Coordination Contract, the Crosschain Transaction is ready to be committed on all sidechains. The Coordinating Node on the Originating Sidechain sends a message requesting that all nodes check the Crosschain Coordination Contract. 
\item Each validator checks the Crosschain Coordination Contract to see if the Crosschain Transaction was committed or ignored. The Coordinating Node applies the state updates if the transaction was committed. If the state is \texttt{Committed} or \texttt{Ignored} the node unlocks the contract.
\end{enumerate}

\subsection{Crosschain View Processing}
Crosschain Views are view calls which call across sidechains within a Multichain Node. Whereas Crosschain Transactions have at their top level an Originating Transaction, Crosschain Views have at their top level a view call. As the view calls only read state, no interaction with other nodes on any of the sidechains is required. As the value of results of each view call on each sidechain does not need to be proven to validators other than the Coordinating Node, the results do not need to be signed. As there are no updates, no contracts need to be locked. The Crosschain View Processing can occur synchronously. 

As Crosschain View calls do not update state based on values read from the distributed ledger, there is flexibility with how locked contracts are treated. The Crosschain View calls can:
\begin{itemize}
\item Fail if any contract called is locked.
\item If any contract is locked, process the call assuming the lock fails.
\item If any contract is locked, process the call assuming the lock succeeds.
\end{itemize}

\section{Programming Model}
\label{section:programming-model}
This section provides some guidelines and considerations when designing applications which initiate Atomic Crosschain Transactions.

\subsection{Transactions and Views}
In the existing Ethereum programming model \cite{programmers-guide-serpent}, Ethereum Transactions are asynchronous \cite{sync-async1996}\cite{distributed-consensus1985}\cite{asynchronous-programming-model2009}. They are executed some time after submission and execution may or may not occur. Additionally, Ethereum Transactions can not return a value. However, a transaction hash is returned to allow the status of the transaction to be tracked. The programming model required to support Crosschain Transactions is similarly asynchronous.

Ethereum Views currently execute synchronously. They execute immediately on the Ethereum node the request is submitted to on a local copy of the distributed ledger. Crosschain Views will similarly execute synchronously on the  Multichain Node the request is submitted to on local copies of the distributed ledgers.

\subsection{Designing Contracts for Locking}
When a Crosschain Transaction fails because it is trying to lock a contract which is already locked, the contract is said to be in lock contention. To minimise lock contention, contracts should be designed to handle only small amounts of data. This is in contrast with the current trend to have large monolithic contracts which manage a lot of data. An example is shown in the Section \ref{section:atomicexample}, \textit{Atomic Swap Ether Transfer Example}, which explains this in detail.

\subsection{Call Depth}
This technology allows for arbitrarily deep call depths between sidechains. That is, there is no hard limit to the nesting of Subordinate Transactions and Subordinate Views below an Originating Transaction. The deeper the call depth due of a Crosschain Transaction, the more likely a call is to encounter a contract which is locked and fail. We are actively researching guidelines for call depth (see Section \ref{section:futurework}, \textit{Future Work}). While we are developing these guidelines, we suggest a call depth of two or three. That is, limiting Crosschain Transactions to an Originating Transaction calling a Subordinate Transaction or View, which calls a Subordinate Transaction or View. 

\subsection{Atomic Swap Ether Transfer Example}
\label{section:atomicexample}
Imagine contracts which facilitate atomic swaps of Ether between Sidechains A and B. On each sidechain there is an Atomic Swap Registration Contract. These contracts are Nonlockable Contracts. When an entity wishes to offer Ether on Sidechain A for Ether on Sidechain B, it deploys a new Atomic Swap Execution Contract on each sidechain. These contracts would be Lockable Contracts. The execution contracts indicate how much Ether the entity has available on Sidechain A and what exchange rate it is prepared to offer. The entity registers each of the execution contracts with the registration contracts on each sidechain. A second entity wishing to offer Ether on Sidechain B for Ether on Sidechain A could monitor the Atomic Swap Registration Contracts, executing repeated Crosschain View calls. The Crosschain View calls could check for matching Atomic Swap Execution Contracts which are deployed on each sidechain, which offer Ether on sidechain A at an acceptable exchange rate. The Crosschain View call could return the address of the Atomic Swap Execution Contract on Sidechain A. The second entity could then execute a Crosschain Transaction to affect the swap, executing the transaction against the Atomic Swap Execution Contract on Sidechain A. The Atomic Swap Execution Contract on Sidechain A would call the Atomic Swap Execution Contract on Sidechain B.

An important characteristic of the atomic swap technique described in the previous paragraph is that the second entity does not need to swap for all of the value which the first entity is offering. In previous techniques  \cite{hashtimelock} the amount being swapped had to be the entire amount. Requiring the entire amount be swapped is a major limitation as this requires the exchanging entities to agree on the amount to be swapped out of band prior to executing the atomic swap on the blockchain. 

No entity can lock the contract thus blocking the registration of new atomic swaps because the Atomic Swap Registration Contract is an Nonlockable Contract. As the Atomic Swap Execution Contract is a Lockable Contract, entities can be sure that swaps will occur atomically. They can be sure that only one entity is able execute an atomic swap on the Atomic Swap Execution Contract at a time. However, many Atomic Swap Execution Contracts can be registered with the Atomic Swap Registration Contract, thus allowing a multitude of atomic swaps to occur simultaneously.

\section{Failure Cases Handled Within Protocol}
This section describes how failures such as nodes going off-line or network connections breaking are handled by the protocol. In particular, all failures and potential failures described in this section are handled automatically by the protocol, and do not require application intervention. 

\subsection{Multichain Node not on all Sidechains}
\underline{Situation}: Multichain Nodes must have validator nodes on all sidechains which are called in the Crosschain Transaction. When the Originating Transaction is submitted to the Coordinating Node on the Originating Sidechain, the Coordinating Node has to assess whether it has all of the sidechains available given the Subordinate Transactions and Subordinate Views to be processed.

\underline{Issue}: If a Coordinating Node does not think that the Multichain Node has access to all of the sidechains necessary to complete the Crosschain Transaction, then it should abort the Crosschain Transaction as soon as possible, to prevent resource wastage.

\underline{Action}: Coordinating Nodes return an error to the caller if the tree of Subordinate Transactions and Subordinate Views contains a transaction or view which needs to be submitted to a sidechain which is not part of the Multichain Node.

\subsection{Common Signer for Subordinate Views and Transactions}
\underline{Situation}: When a Coordination Node receives a tree of Subordinate Transactions and Subordinate Views, they may not all be signed by the same account.

\underline{Issue}: Allowing Subordinate Transactions or Subordinate Views which were signed by accounts other than the one submitting the top level Transaction or signed View could possibly be used as some form of replay attack. 

\underline{Action}: An error is returned to the application.

\subsection{Crosschain Transaction Replay}
\underline{Situation}: An attacker may have a copy of an old Crosschain Transaction. The transaction may have failed to be committed. The attacker should not be able to replay the transaction. 

\underline{Issue}: If a Crosschain Transaction could be replayed, attackers could attempt to attack the system with the replayed transaction.

\underline{Action}: The transaction will be rejected by the Crosschain Coordination Contract when the Coordination Node on the Originating Sidechain submits the Crosschain Transaction Start message because an entry will already exist for the combination of Originating Sidechain Identifier and Crosschain Transaction Identifier on the Coordination Blockchain specified in the transaction.

\subsection{Coordinating Node on the Originating Blockchain Failures}
\label{section:cnos}
\underline{Situation 1}: The Coordinating Node on the Originating Blockchain fails before the Crosschain Transaction Start message has been accepted by the Crosschain Coordination Contract.

\underline{Issue 1}: The Crosschain Transaction will not start. 

\underline{Action 1}: The Crosschain Transaction will not be committed.

\underline{Situation 2}: The Coordinating Node on the Originating Blockchain fails at any point before the Crosschain Transaction Commit message has been accepted by the Crosschain Coordination Contract.

\underline{Issue 2}: The Crosschain Transaction will time-out, and then be ignored.

\underline{Action 2}: The Crosschain Transaction will not be committed.

\underline{Situation 3}: The Coordinating Node on the Originating Blockchain fails after the Crosschain Transaction Commit message has been accepted by the Crosschain Coordination Contract, but before it has sent out the request for all nodes to look at the Crosschain Coordination Contract.

\underline{Issue 3}: Nodes will not know that the Crosschain Transaction has been committed.

\underline{Action 3}: All of the nodes will have set-up a local timer which should expire when the Transaction Timeout Block Number is exceeded. Alternatively, they could wait until they need to access the contract state. Either way, they check the Crosschain Coordination Contract to see if the Crosschain Transaction should be committed or ignored.

\subsection{Coordinating Node on a Sidechain Executing a Subordinate View or Transaction Fails}
\label{section:cns}
\underline{Situation 1}: The Coordinating Node on a sidechain executing a Subordinate Transaction or View fails before the Subordinate Transaction or View has been submitted to the sidechain.

\underline{Issue 1}: The Coordinating Node will not be available to process the Subordinate Transaction or View.

\underline{Action 1}: The Coordinating Node on the sidechain which submitted the Subordinate Transaction or View will not be able to contact the Coordinating Node. If it is processing a Subordinate View, it will propagate an error to its caller. If it is processing a Subordinate Transaction, it will propagate an error to the Coordinating Node on the Originating Sidechain. The Coordinating Node on the Originating Sidechain will work with other nodes on the Originating Sidechain to sign a Crosschain Transaction Ignore message, and will submit it to the Crosschain Coordination Contract. The Crosschain Transaction will be ignored.

\underline{Situation 2}: The Coordinating Node on a sidechain executing a Subordinate View fails after the Subordinate View has been submitted but prior to the Subordinate  View Result message being returned. 

\underline{Issue 2}: The Coordinating Node which submitted the Subordinate View will not get the result of the Subordinate View.

\underline{Action 2}: The Coordinating Node on the sidechain which submitted the Subordinate View will time-out waiting for the result. It will propagate an error to its caller in the case of a Subordinate View or to the Coordinating Node on the Originating Sidechain in the case of a Subordinate Transaction. The Coordinating Node on the Originating Sidechain will work with other nodes on the Originating Sidechain to sign a Crosschain Transaction Ignore message, and will submit it to the Crosschain Coordination Contract. The Crosschain Transaction will be ignored. 

\underline{Situation 3}: The Coordinating Node on a sidechain executing a Subordinate Transaction fails after the Subordinate Transaction has been submitted and mined, and the Subordinate Transaction Ready message has been threshold signed, but prior to the Subordinate Transaction Ready message being sent to the Coordinating Node on the Originating Sidechain. 

\underline{Issue 3}: The Coordinating Node on the Originating Sidechain will not receive a Subordinate Transaction Ready message for the Subordinate Transaction. However, the nodes on the sidechain have mined the Subordinate Transaction.

\underline{Action 3}: The Coordinating Node on the Originating Sidechain will time-out waiting for the Subordinate Transaction Ready message. It will work with other nodes on the Originating Sidechain to sign a Crosschain Transaction Ignore message, and will submit it to the Crosschain Coordination Contract. The nodes on the sidechain with the failed Coordination Node will have set-up a timer which should expire when the Transaction Timeout Block Number is exceeded. Alternatively, they could wait until they need to access the contract state. Either way, they check the Crosschain Coordination Contract to see if the Crosschain Transaction should be committed or ignored. The Crosschain Transaction will be ignored.

\underline{Situation 4}: The Coordinating Node on a sidechain executing a Subordinate Transaction or View fails before it receives the request to check the Crosschain Coordination Contract for the commit state.

\underline{Issue 4}: Nodes on the sidechain will not know that the Crosschain Transaction has been committed.

\underline{Action 4}: All of the nodes on the sidechain will have set-up a timer which should expire when the Transaction Timeout Block Number is exceeded. Alternatively, they could wait until they need to access the contract state. Either way, they check the Crosschain Coordination Contract to see if the Crosschain Transaction should be committed or ignored.

\subsection{General Node Failures}
\label{section:gnf}
\underline{Situation}: A node which isn't a Coordinating Node fails.

\underline{Issue}: If enough nodes fail, the sidechain will not be able to threshold sign messages.

\underline{Action}: If any of the threshold signed messages can not be signed, the transaction will fail. If the Crosschain Transaction Ignore message can not be created, then the transaction will time-out. The Crosschain Transaction will be ignored. 

\subsection{Nodes Removed from a Sidechain}
\underline{Situation}: Nodes can be removed from sidechain due to voting or other sidechain specific mechanisms. If this occurs, then the node will no longer have permission to connect to other nodes on the sidechain.

\underline{Issues and Actions}: This is an equivalent situation to a node failing. See Sections \ref{section:cnos}, \ref{section:cns}, and \ref{section:gnf} for details.

\subsection{Network Connection Failures}
\underline{Situation}: Network connections between nodes could fail. This is less likely to occur between nodes in a Multichain Node than for nodes in a sidechain, as nodes in a Multichain Node are likely to be co-located and nodes in a sidechain are likely to be widely dispersed.

\underline{Issues and Actions}: Network failure scenarios will be similar to a node failing. See Sections \ref{section:cnos}, \ref{section:cns}, and \ref{section:gnf} for details.

\subsection{Attacker Compromises Nodes}
\underline{Situation}: An attacker could fully compromise a node. They could execute arbitrary code on the node. 

\underline{Issue 1}: The attacker could try to create Crosschain Transaction Start messages with long time-outs, hoping to lock contracts for long periods. They could try to create Subordinate View Result messages with incorrect results. They could try to create Crosschain Transaction Commit messages to attempt to commit failed Crosschain Transactions.

\underline{Action 1}: Nodes will not threshold sign messages which they do not agree with. The resulting action will be for the Crosschain Transaction to fail.

\underline{Issue 2}: If the attacker was a Coordinating Node they could hold a Subordinate Transaction. If the application resubmits the transaction (see Section \ref{section:resubmitting}, \textit{Resubmitting Failed Crosschain Transactions}) prior to the original transaction timing-out, the attacker could release the Subordinate Transaction and have it mined ahead of the resubmitted Subordinate Transaction. The resubmitted Subordinate Transaction would fail as it would use a \textit{nonce} value which matched the \textit{nonce} value used by the released transaction. This could result in a scenario similar to Section \ref{section:livelock}, \textit{Livelock}.

\underline{Action 2}: The application should wait for the original Crosschain Transaction to time-out prior to submitting new transactions. The attacker behaviour could be detected. The compromised node could be removed from the sidechain while the owner of the Multichain Node containing the compromised node removed the attacker.

\subsection{Threshold Signing Issues}
\label{section:thresholdsigningissues}
\underline{Situation}: Coordinating Nodes create messages which need to be threshold signed in various parts of the protocol. Not enough of the validators may cooperate to create the threshold signature. This could happen for several reasons including: some validators do not believe the message should be signed, some validators being offline or some validators are uncontactable. Validators will send error message to the Coordinating Node if they do not want to sign the message. The Coordinating Node sets a local timer to detect time-out conditions whilst waiting for partial signatures. Independent of the reason, the Coordinating Node will be left not able to complete the threshold signing of the message.

\underline{Issue}: The issue depends on the message which can not be signed:
\begin{itemize}
\item Crosschain Transaction Start: The Coordinating Node on the Originating Sidechain can not start the Crosschain Transaction by submitting the Crosschain Transaction Start message to the Crosschain Coordination Contract.
\item Crosschain Transaction Commit: The Coordinating Node on the Originating Sidechain can not commit the Crosschain Transaction by submitting the Crosschain Transaction Commit message to the Crosschain Coordination Contract.
\item Crosschain Transaction Ignore: The Coordinating Node on the Originating Sidechain can not request the Crosschain Transaction be ignored early, thus unlocking locked contracts early, by submitting the Crosschain Transaction Ignore message to the Crosschain Coordination Contract.
\item Subordinate Transaction Ready: The Coordination Node on a sidechain can not indicate to the Coordinating Node on the Originating Sidechain that the Subordinate Transaction has been mined and is ready to be committed. 
\item Subordinate View Result:  The Coordination Node on a sidechain can not return a Subordinate View result to the Coordinating Node which instigated the call. 
\end{itemize}

\underline{Action}: The action depends on the message which can not be signed:
\begin{itemize}
\item Crosschain Transaction Start: The Coordinating Node on the Originating Sidechain terminates the Crosschain Transaction.
\item Crosschain Transaction Commit: The Crosschain Transaction will time-out when the block number on the Coordination Blockchain is greater than the Transaction Timeout Block Number.
\item Crosschain Transaction Ignore: The Crosschain Transaction will time-out when the block number on the Coordination Blockchain is greater than the Transaction Timeout Block Number.
\item Subordinate Transaction Ready: The Coordination Node on the sidechain will indicate to the Coordinating Node on the Originating Sidechain an error indicating that the Subordinate Transaction Ready message can not be created. The Coordinating Node on the Originating Sidechain will attempt to create a Crosschain Transaction Ignore message to terminate the transaction early. 
\item Subordinate View Result: An error message is returned to the calling Coordinating Node. The error is propagated up to the Coordinating Node on the Originating Sidechain. The Coordinating Node on the Originating Sidechain will attempt to create a Crosschain Transaction Ignore message to terminate the transaction early. 
\end{itemize}

\subsection{Coordination Blockchain Congestion}
\underline{Situation}: The Coordinating Node on the Originating Sidechain submits an Ethereum Transaction to start, commit or ignore the Crosschain Transaction to the Crosschain Coordination Contract on the Coordination Blockchain. If there is a Denial of Service (DoS) attack on the blockchain nodes, or if the blockchain utilisation is very high, the transaction may not be accepted.

\underline{Issue}: The issues are the same as is described in Section \ref{section:thresholdsigningissues}, \textit{Threshold Signing Issues}.

\underline{Action}: The actions are the same as is described in Section \ref{section:thresholdsigningissues}, \textit{Threshold Signing Issues}.

\subsection{Sidechain Public Key Issues}
\underline{Situation}: Threshold signed messages are generated by validator nodes on sidechains to prove values to entities not on the sidechain. The messages are verified using Sidechain Public Keys which can be obtained from the Coordination Blockchain. The signature on a message could fail to verify when a validator was added or removed from the sidechain, during the transition time when the public key was updated on the Coordination Blockchain. 

\underline{Issue}: Crosschain Transactions will fail if threshold signatures can not be verified.

\underline{Action}: The validator nodes should continue to use old threshold private keys until the new public key is available on the Coordination Blockchain. The old public key should be available and able to be used for a short period after the new public key is published, to allow threshold messages which have been signed but not verified to be verified.

\subsection{Adding Nodes to a Sidechain}
\underline{Situation}: When a new sidechain node joins the network, it will attempt to synchronise the blockchain.

\underline{Issue 1}: Based solely on the blockchain data, the new node will not know which Originating and Subordinate Transactions should be committed and which should be ignored.

\underline{Action 1}: For each Originating and Subordinate Transaction, the node needs to check with the appropriate Coordination Blockchain and Crosschain Coordination Contract to see if the transaction was committed or not.

\underline{Issue 2}: The new node may not be able to reach the appropriate Coordination Blockchains, which could stop its ability to synchronise the blockchain.

\underline{Action 2}: There is no mitigation for this situation. It may be wise to periodically (or on another basis) create a new genesis block which integrates the current state of the distributed ledger for long-lived sidechains to limit the potential impact. This would be similar to W3C's \textit{Checkpoint Block} \cite{w3c-blockchain}.

\subsection{Crosschain Transaction with Existing Contracts}
\underline{Situation}: An existing blockchain system may receive an upgrade which enables Crosschain Transactions.

\underline{Issue}: Contracts on the existing blockchain may not have been written to cater for Crosschain Transactions. In particular, the locking caused by a Crosschain Transaction may adversely affect the system. 

\underline{Action}: Contracts which have not been created with a \texttt{Lockable} flag can not be locked. As such, all contracts on the existing system would not be lockable. Attempting to perform a Crosschain Transaction involving these contracts would fail as the transaction would fail when it attempted to gain a lock on the contract.

\subsection{Livelock}
\label{section:livelock}
\underline{Situation}: Suppose that the contracts in Listings \ref{listing:livelock1} and \ref{listing:livelock2} are deployed on Sidechains 1 and 2 respectively. They each execute two Crosschain Transactions, one calling a function in a contract on Sidechain 3 first and then a function in a contract on Sidechain 4, and the other in the opposite order. 
\begin{lstlisting}[
  frame=single,
  basicstyle=\footnotesize\ttfamily,
  numbers=left,
stepnumber=1, 
  firstnumber=1,
  numberfirstline=true,
  numbersep=5pt,    
  xleftmargin=0.5cm,
  morekeywords={contract, function},
  caption={Livelock Failure Case: Sidechain 1},
  captionpos=b,
  label={listing:livelock1},
]
contract Contract1 {
  function foo() {
    sc3.c31.buy()
    sc4.c41.sell()
  }
}
\end{lstlisting}

\begin{lstlisting}[
  frame=single,
  basicstyle=\footnotesize\ttfamily,
  numbers=left,
stepnumber=1, 
  firstnumber=1,
  numberfirstline=true,
  numbersep=5pt,    
  xleftmargin=0.5cm,
  morekeywords={contract, function},
  caption={Livelock Failure Case: Sidechain 2},
  captionpos=b,
  label={listing:livelock2},
]
contract Contract2 {
  function bar() {
    sc4.c41.sell()
    sc3.c31.buy()
  }
}
\end{lstlisting}

The following sequence of events may occur:
\begin{itemize}
\item Application A submits a Crosschain Transaction to Multichain Node A for \texttt{Contract1.foo()} on Sidechain 1.
\item Application B submits a Crosschain Transaction to Multichain Node B for \texttt{Contract2.bar()} on Sidechain 2.
\item The Crosschain Transaction Start messages for the Crosschain Transactions are signed and accepted by a Crosschain Coordination Contract for Sidechain 1 and Sidechain 2.
\item Multichain Node A gets Sidechain 3 to mine the Subordinate Transaction which is calling \texttt{c31.buy()}, that locks contract \texttt{c31}.
\item Multichain Node B gets Sidechain 4 to mine the Subordinate Transaction which is calling \texttt{c41.sell()}, that locks contract \texttt{c41}.
\item The Crosschain Transaction initiated by Application A now fails when executing \texttt{sc4.c41.sell()} because contract \texttt{c41} is locked.
\item The Crosschain Transaction initiated by Application B now fails when executing \texttt{sc3.c31.buy()} because contract \texttt{c31} is locked.
\item Both transaction fails.
\item Both applications are notified of the failed transactions.
\item Both applications will submit the respective transactions again.
\item This situation could happen in perpetuity.
\end{itemize}

\underline{Issue}: The sequence of events outlined above leads to a \textit{livelock} situation where neither the transaction \texttt{c1.foo()} nor the transaction \texttt{c2.bar()} will ever be committed.

\underline{Action}: There is no solution for the moment. However, it should be noted that the probability that none of the transactions will be committed decreases as time progresses.

\subsection{Centralisation}
\underline{Situation}: It could be considered that this proposed approach is centralised due to the existence of concepts such as Originating Blockchain and Coordination Nodes.

\underline{Issue}: Centralisation is against the ethos of blockchain.

\underline{Action}: The Crosschain Transaction system has per transaction centralisation. That is, there are single Coordinating Nodes per sidechain, and in particular there is a single Coordinating Node on the Originating Sidechain. If these nodes or the connections between them fail, the crosschain transaction fails. Despite these per transaction centralisation points, the overall system is decentralised. Any Multichain Node can instigate a crosschain transaction from any sidechain.

\section{Failure Cases Handled By Application}
This section explains how application design can help prevent failures such as a contract being continually locked. In particular, all failures and potential failures described in this section need to be handled by the application, because they are not automatically handled by the protocol. 

\subsection{Heavily Used Contracts Continually Being Locked}
\underline{Situation}: Contracts used by many entities could be continually locked.

\underline{Issue}: Entities wishing to read data from the contract could not read data. The contract could become unusable.

\underline{Action}: Follow the guidelines in Section \ref{section:programming-model}, \textit{Programming Model}.

\subsection{Resubmitting Failed Crosschain Transactions}
\label{section:resubmitting}
\underline{Situation}: Crosschain Transactions may fail for a variety of reasons. The Crosschain Transaction needs to be resubmitted. 

\underline{Issue}: None, some or all of the Originating Transaction and Subordinate Transactions may have been mined. Ethereum Transactions require the next Ethereum Transaction Nonce must be used. Ethereum Clients reject transactions submitted with out of order nonces.

\underline{Action}: A new Crosschain Transaction must be created as the Crosschain Transaction Identifier must be unique. Prior to creating the new Crosschain Transaction, the application must determine what nonce values should be used for the Originating Transaction and Subordinate Transactions, given some of the Originating Transaction and Subordinate Transactions may have been mined in the failed Crosschain Transaction.

\subsection{Scaling within a Sidechain}
\underline{Situation}: The threshold signature scheme used by this approach requires \texttt{2N} messages to be passed between nodes on a sidechain for each threshold signed message. A Crosschain Transaction requires two threshold signed messages (Start and Commit or Ignore) plus one threshold signed message for each Subordinate Transaction and for each Subordinate View. 

\underline{Issue}: Ethereum Private Sidechains using the consensus algorithms in the IBFT family (e.g. Clique, IBFT, IBFT 2) will encounter practical limitations in the number of validators due to the amount of message traffic generated by the protocols. Additionally, sidechains using IBFT may also encounter security vulnerabilities when operating on partially synchronous networks \cite{ibft-correctness}.

\underline{Action}: Limit the number of nodes in a sidechain which is likely to be part of Crosschain Transactions. The research team plan to determine quantifiable limits as described in Section \ref{section:futurework}, \textit{Future Work}.

\subsection{Coordination Blockchain Bottleneck}
\underline{Situation}: All Crosschain Transactions require two Ethereum Transactions to be submitted to the Coordination Blockchain; one to start the transaction and one to commit or ignore the transaction. If many sidechains use the one Coordination Blockchain, then it could become overwhelmed, and start and commit messages may not be able to be submitted when needed.

\underline{Issue}: If a Coordinating Node on an Originating Sidechain can not successfully submit a Crosschain Transaction Start message then the Crosschain Transaction can not be started and will fail. If the Crosschain Transaction Commit message can not be successfully submitted, then a transaction which could have been committed will time-out and be ignored.

\underline{Action}: A multitude of Coordination Blockchains could be used. These blockchains could in fact be Ethereum Private Sidechains. Different Coordination Blockchains could be used with each Crosschain Transaction. However, all nodes which will be involved in the transaction need to be able to access the Coordination Blockchain.

\subsection{Coordination Blockchain Availability}
\underline{Situation}: When a new node joins a sidechain, it needs to access all Coordination Blockchains referenced in Originating and Subordinate Transactions to determine if the transactions in the blockchain should be committed or ignored.

\underline{Issue}: This implies that Coordination Blockchains need to live for as long as the oldest sidechain which they are referenced in. 

\underline{Action}: Prior to archiving any Coordination Blockchain, audit all operational sidechains to ensure no transaction refers to the Coordination Blockchain. An alternative to this is from time to time create an authenticated snapshot of the blockchain state, which new nodes can use as a starting point for synchronising the blockchain. This would be similar to W3C's \textit{Checkpoint Block} \cite{w3c-blockchain}.

\subsection{Privacy of Transactions in Crosschain Coordination Contract}
\underline{Situation}: The Coordinating Node on the Originating Sidechain submits an Ethereum Transactions to start, commit and ignore Crosschain Transactions. These transactions include the Originating Sidechain Identifier. These transactions must be signed by an Ethereum Account on the Coordination Blockchain.

\underline{Issue}: The rate of Crosschain Transactions originating from a particular sidechain for a particular enterprise is revealed on the Coordination Contract. The Ethereum Account on the Coordination Blockchain might be able to be linked to a specific enterprise. If this could be done, then the fact that a specific enterprise is a member of the sidechain would be revealed.

\underline{Action}: There is no mitigation for this situation.

\subsection{Adding and Removing Nodes from a Sidechain}
\underline{Situation}: The Crosschain Transaction System described in this paper relies on threshold signatures for proving values across sidechains, and threshold voting for updating Sidechain Public Keys in the Crosschain Coordination Contract.

\underline{Issue}: Adding a validating node to a sidechain and not adjusting the threshold will make it easier for validators to collude to produce malicious threshold signed messages or change the Sidechain Public Key. Removing a validating node from a sidechain and not adjusting the threshold will make it harder and could make it impossible to sign messages.

\underline{Action}: As validating nodes are added and removed from a sidechain, the threshold used for distributed key generation and for voting on Sidechain Public Keys needs to be adjusted. The threshold could be set to match the threshold of the consensus algorithm. In this way, if a block can be produced on a sidechain, then so too will a threshold signed message be able to be produced.

\subsection{Trust Boundaries}
\label{section:trust}
\underline{Situation}: On a single blockchain, the trust boundary is defined around the entire chain including all of its nodes, but excluding applications such as wallets that might connect to it. The trust boundary for a single chain is guaranteed by the nature of the chain itself and is therefore both implemented and maintained at a technical level. Crosschain transactions introduce an expansion of the trust boundary that must extend a priori to all other sidechains and coordination blockchains that are utilised by the Originating Sidechain. 

\underline{Issue}: The crosschain trust boundary cannot be defined statically because it may be extended at runtime by authors of smart contracts. The trust boundary also cannot be enforced technically. Therefore, crosschain trust boundaries must be both defined and managed socially, at a level above the sidechain itself.

\underline{Action}: Originating Sidechain operators must choose to trust operators and implementations of all other sidechains and Coordination Blockchains that are utilised by the Originating Transaction, or restrict smart contracts from being able to access them via out of band mechanisms; for example white or black network connection lists or firewall rules. Setting only selected smart contracts as Lockable will act to reduce the Trust Boundary.

\subsection{Byzantine Coordination Blockchains}
\underline{Situation}: Byzantine Coordination Blockchains could negatively impact state on Originating Sidechains by selectively answering checks from other validator nodes on whether Crosschain Transactions have been committed. 

\underline{Issue}: Such behaviour could result in forking validators on Originating Sidechains, and could conceivably place them in states from which they cannot recover the correct state of their blockchain.

\underline{Action}: The actions are the same as is described in Section \ref{section:trust}, \textit{Trust Boundaries}.

\subsection{Liveness Issue 1}
\underline{Situation}: The overall combined minimum network latency is greater than the time-out specified by the Transaction Timeout Block Number.

\underline{Issue}: The Crosschain Transaction cannot be executed as the transaction will always time-out. The block number on the Coordination Blockchain will always be greater than the Transaction Timeout Block Number before a Crosschain Transaction Commit message can be submitted to the Coordination Blockchain.

\underline{Action}: The Application should submit a new Crosschain Transaction with a longer timeout. Note that the guidelines in Section \ref{section:resubmitting}, \textit{Resubmitting Failed Crosschain Transactions} should be taken into consideration prior to submitting the new transaction. 

\subsection{Liveness Issue 2}
\underline{Situation}: The minimum network latency is higher than the largest Crosschain Transaction Time-out the Crosschain Coordination Contract is configured to allow.

\underline{Issue}: The Crosschain Transaction cannot be executed as the transaction will always time-out. The block number on the Coordination Blockchain will always be greater than the Transaction Timeout Block Number before a Crosschain Transaction Commit message can be submitted to the Coordination Blockchain.

\underline{Action}: Use a Crosschain Coordination Contract which is configured to allow larger Crosschain Transaction Time-outs.

\subsection{DoS using Subordinate Reads}
\underline{Situation}: Imagine there were three sidechains A, B, and C, that Enterprise A has a Multichain Node on all three sidechains, and Enterprise B has a Multichain Node on sidechains B and C. Further imagine that there is a time sensitive bidding process on sidechain C which Enterprise A and B are competing on. To lodge a bid, Enterprise A must do a Crosschain Transaction involving sidechains A and C, and Enterprise B must do a Crosschain Transaction involving B and C. Enterprise A might be able to execute many legitimate Crosschain Transactions which execute Subordinate Views on sidechain B.

\underline{Issue}: Enterprise A might be able to overload the computing resources for sidechain B, thus blocking Enterprise B's ability to submit a bid.

\underline{Action}: The actions are the same as is described in Section \ref{section:trust}, \textit{Trust Boundaries}.

\section{Applying Technology to Other Blockchains}
\subsection{Private Blockchains}
The atomic crosschain transaction technology described in this paper could be applied to a wide variety of private blockchain and distributed ledger systems including Hyperledger Fabric \cite{hyperledger-fabric}\cite{hyperledger-fabric-doc}, Corda \cite{corda-doc} and Quorum \cite{quorum-source}\cite{quorum-wiki}\cite{quorum-examples}. To operate as a sidechain within a Multichain Node, using an Ethereum blockchain for the Coordination Blockchain, the platforms would need to offer the following features:
\begin{itemize}
\item \textbf{Validators}: A set of validators which can attests to the state of the blockchain. 
\item \textbf{Threshold signature scheme and associated distributed key generation and messaging}: This is required so that the blockchain can generate crosschain messages such as the Subordinate View Result message.
\item \textbf{Finality}: The consensus algorithm used on the blockchain must be able to determine when a block is final. This is required because crosschain messages such as the Crosschain Transaction Ready and the Crosschain Transaction Commit messages can not be issued until the Subordinate and Originating Transactions they relate to are final.
\item \textbf{Contract locking and unlocking mechanisms}: This is required so that transactions across blockchains will be atomic.
\item \textbf{Subordinate View / Transaction Execution}: A platform specific method for indicating a Subordinate Transaction or View should be executed within the contract code. In Ethereum platforms this takes the form of a precompile. 
\item \textbf{RLP encoded function calls}: The encoding format for crosschain transactions is RLP encoding. As such, function call parameter values need to be translated to and from the native blockchain format to RLP encoded values. 
\item \textbf{JSON RPC \texttt{cc\_SendRawCrosschainTransaction} API method}: This is the API called by other blockchains to execute Subordinate Views and Subordinate Transactions.
\end{itemize}

\subsection{Public Blockchains}
The underlying assumption of the crosschain transaction technology is that all participants are known and wish to maintain their reputation. It is assumed that the consequences for acting inappropriately occur outside of the blockchain. Having this assumption allows a set of validators to threshold sign a message which attests to the state of the blockchain. If a permissionless blockchain system could meet this assumption, then they too could support this technology.

\section{Conclusion}
We introduced a methodology for Atomic Crosschain Transactions across private sidechains and blockchains. Using the techniques described in this paper, a function in a contract on one sidechain could call functions in contracts on other sidechains. This capability allows functionality or data available in one sidechain to be used by other sidechains, which opens up a swathe of new opportunities within the blockchain ecosystem. The onus rests on the application designers and developers to adhere to the programming model and guidelines to fully harness this powerful technology, and in the spirit of the open source community contribute to extending this functionality to make it even more robust and flexible.

This technology has been described in terms of Ethereum Private Sidechains. However, it could be used with any contract based private blockchain system which incorporates the requisite features. In particular, the ideas in this paper could facilitate an inter-ledger protocol which would allow disparate blockchain technologies to interoperate.

\section{Future Work}
\label{section:futurework}
The following areas are being actively pursued as part of the Crosschain Transactions work.

\begin{itemize}
\item Theoretical and practical performance analysis. In particular, to determine the performance of Crosschain Transactions as the number of nodes in a sidechain grows and the Subordinate Transaction and View depth grows.
\item Investigate more complex locking mechanisms and the potential to invoke different types, or levels, of locking depending on the call graph and complexity of the Crosschain Transactions.
\item Statistical analysis of the probability of transaction failure due to locking errors based on the call depth of Subordinate Transactions and Views.
\item Charging Gas and mining Ethereum Views to provide an economic cost of calling Ethereum Views.
\item Determine the feasibility and rationale for charging Gas for Subordinate Transactions even if the Crosschain Transaction fails.
\item Formal analysis of the protocol.
\item Crypto economic incentivisation analysis.
\item Programming templates and schemes.
\item Evaluation of  the feasibility of crosschain operations when one of the subordinate chains uses a consensus algorithm with probabilistic finality. 
\end{itemize}

\ifCLASSOPTIONcompsoc
  \section*{Acknowledgments}
\else
  \section*{Acknowledgment}
\fi
This research has been undertaken whilst we have been employed full-time at ConsenSys. Peter Robinson acknowledges the support of University of Queensland where he is completing his PhD, and in particular the support of his PhD supervisor Dr Marius Portmann.

We acknowledge Dr Shahan Khatchadourian and Horacio Mijail Anton Quiles for their input regarding  locking and atomicity, Dr Shahan Khatchadourian for his advice on the authentication of Crosschain Transaction Commit messages posted to the Crosschain Coordination Contract, and Dr Marius Portmann for his contribution to the conversations, the outcome of which was the concept of Crosschain Coordination Contract to hold Crosschain Transaction state and and for his input into discussions on failure cases. 

We thank Dr Catherine Jones, Horacio Mijail Anton Quiles, and Martin Bosticky for reviewing this paper and providing astute feedback.



\bibliographystyle{IEEEtran}
\bibliography{IEEEabrv,ref}
%
%
%

\end{document}